\documentclass[%
reprint,
 amsmath,amssymb,
 aps,
prb,
]{revtex4-2}

\usepackage{graphicx}
\usepackage{dcolumn}
\usepackage{bm}
\usepackage{mathrsfs}
\usepackage{braket}
\usepackage{color}
\usepackage{comment}
\usepackage{siunitx}
\usepackage[FIGTOPCAP,nooneline]{subfigure}
\subcapraggedrighttrue
\usepackage[version=3]{mhchem}

\usepackage{hyperref}
\hypersetup{
     colorlinks   = true,
     linkcolor    = blue,
     citecolor    = blue,
     urlcolor     = blue
}

\begin{document}

\preprint{APS/123-QED}

\title{Thermodynamic formulation of the spin magnetic octupole moment in bulk crystals}

\author{Jun {\=O}ik\'{e}}
\altaffiliation[oike.jun.32y@st.kyoto-u.ac.jp]{}
\author{Robert Peters}
\author{Koki Shinada}
\altaffiliation[koki.shinada@riken.jp \\ Present address: RIKEN Center for Emergent Matter Science (CEMS), Wako, Saitama 351-0198, Japan]{}
\affiliation{%
Department of Physics, Kyoto University, Kyoto 606-8502, Japan}%

\date{\today}

\begin{abstract}
The discovery of unconventional antiferromagnets, such as altermagnets, has drawn significant attention to higher-rank magnetic multipoles, particularly magnetic octupoles.
Despite the advances in research, attempts to understand their microscopic properties remain limited due to the unbounded nature of the position operator in bulk crystals.
In this paper, we address this problem by using a well-known thermodynamic approach and derive a formula for the spin magnetic octupole moment (SMOM) that can be used in bulk crystals.
The resulting formula is gauge invariant and satisfies St\v{r}eda formulas that relate the SMOM to the spin magnetoelectric dipole-quadrupole susceptibilities.
Furthermore, we apply this formula to several models and examine the fundamental properties of the SMOM.
For example, in $d$-wave altermagnets, the nonrelativistic component of the SMOM, which is independent of spin-orbit coupling, is larger than the relativistic component, which is induced by spin-orbit coupling.
These nonrelativistic SMOMs have the same microscopic origin as the nonrelativistic spin splitting that characterizes $d$-wave altermagnetism.
Moreover, they exhibit a N\'{e}el vector dependence consistent with Landau theory for $d$-wave altermagnetism [Phys. Rev. Lett. \textbf{132}, 176702 (2024)].

\end{abstract}

\maketitle

\section{Introduction}

Higher-rank magnetic multipoles can act as order parameters for exotic phases that go beyond conventional magnetic dipoles.
In particular, magnetic octupoles have gained significant attention since the discovery of unconventional antiferromagnets with ferroic octupole orderings, such as altermagnets~\cite{Bhowal2024-vj,Fernandes2023-cw,McClarty2024-eo}, \ce{Mn3Sn}~\cite{Suzuki2017-ve}, and all-in/all-out pyrochlore iridates \ce{$R$2Ir2O7} ($R$=rare earth)~\cite{Arima2013-mp}.
Altermagnets share characteristics of both ferromagnets and antiferromagnets: a ferromagnet-like spin splitting in momentum space and an antiferromagnet-like vanishing net magnetization~\cite{Noda2016-wt,Naka2019-uk,Ahn2019-cy,Hayami2019-zk,Naka2020-pm,Smejkal2020-zu,Yuan2020-al,Yuan2021-xz,Mazin2021-ea,Egorov2021-uo,Smejkal2022-xr,Smejkal2022-xx}. 
Additionally, \ce{Mn3Sn} and all-in/all-out pyrochlore iridates possess Weyl points~\cite{Kuroda2017-oa,Wan2011-sf}, resulting in a sizable anomalous Hall effect~\cite{Nakatsuji2015-ll,Ueda2018-wq}.
These unique characteristics not only expand potential applications of antiferromagnets but also emphasize the significance of their order parameters, which are magnetic octupoles.

Moreover, magnetic octupoles induce various intriguing response phenomena, among which the most fundamental one is the linear piezomagnetic effect~\cite{Bhowal2024-vj,McClarty2024-eo}.
This effect was recently observed in altermagnetic MnTe and enables the control of antiferromagnetic domains in altermagnets~\cite{Aoyama2024-nz}.
Beyond this linear response, their higher-rank nature also activates nonlinear responses such as the nonlinear magnetoelectric effect~\cite{Urru2022-ww,Oike2024-bl,Hu2024-gc} and third-order nonlinear Hall effect~\cite{Fang2023-xl,Sorn2023-be,Farajollahpour2024-yf}.
These nonlinear octupolar responses could realize all-electrically controlled spintronics in antiferromagnets (e.g., altermagnets) that do not mainly exhibit dipolar responses such as the anomalous Hall effect.
Furthermore, the nonequilibrium transport of magnetic octupoles themselves has been theoretically proposed~\cite{Tahir2023-di} and could induce a torque on the N\'{e}el vector of altermagnets~\cite{Han2024-wn}.

To further unlock their potential, it is essential to directly calculate the magnetic octupole moment (MOM) as an order parameter.
However, calculating the MOM in bulk crystals is a challenging problem and is currently limited to atomic-scale calculations~\cite{Bhowal2024-vj,Suzuki2017-ve}. 
The main reason is that the fundamental definition of the MOM operator, $O_{ijk} =\int d\bm{r} \mu_i(\bm{r})r_jr_k $, involves the position operator $\bm{r}$, which is unbounded in bulk crystals. 
Here, $\bm{\mu}(\bm{r})$ denotes the magnetization density, and $i,j,k$ label Cartesian components.
An approach to overcome this difficulty is to redefine the MOM based on a thermodynamic relation.
To date, this prescription has proven effective for formulating other multipole moments that originally include the position operator, such as orbital magnetization~\cite{Shi2007-xs}, magnetic quadrupoles~\cite{Gao2018-uk,Shitade2019-hq,Shitade2018-rd,Gao2018-no}, and electric quadrupoles~\cite{Daido2020-yo}.

In this paper, we derive a gauge-invariant formula of the spin contribution of the MOM, i.e., the spin MOM (SMOM), using a thermodynamic relation.
The physical meaning of the resulting formula is clarified by taking the molecular limit, revealing that it consists of transition rate matrices and other elements from various multipoles.
Furthermore, the SMOM satisfies St\v{r}eda formulas related to the spin magnetoelectric dipole-quadrupole susceptibilities, which describe the response between dipolar and quadrupolar fields in magnetoelectric-coupled regimes. 
In addition, we confirm the validity of the St\v{r}eda formulas through numerical calculations and find that the corresponding response coefficients are enhanced due to interband effects.
Note that we do not consider the orbital contribution here because spin magnetization usually dominates the total magnetization~\cite{Meyer1961-tf,reck1969orbital}.

Furthermore, we examine the microscopic properties of the SMOM in altermagnets with magnetic octupole orders, i.e., $d$-wave altermagnets~\cite{Bhowal2024-vj}.
The SMOM includes a relativistic component originating from spin-orbit coupling (SOC) and a nonrelativistic one independent of it; in $d$-wave altermagnets, the nonrelativistic component is larger than the relativistic one.
These nonrelativistic SMOMs have the same microscopic origin as the nonrelativistic spin splitting that characterizes $d$-wave altermagnetism.
Moreover, they exhibit a N\'{e}el vector dependence consistent with Landau theory for $d$-wave altermagnetism~\cite{McClarty2024-eo}, which shows that they couple linearly to the N\'{e}el vector.

The rest of this paper is organized as follows:
In Sec.~\ref{sec:octupole_derivation}, we derive the SMOM and discuss its physical meaning.
Specifically, we introduce the thermodynamic definition of the SMOM in Sec.~\ref{subsec:octupole_definition} and confirm the gauge invariance of the derived expression in Sec.~\ref{subsec:octupole_expression}.
Section~\ref{subsec:molecular_limit} is devoted to explaining the physical meaning by taking the molecular limit, and Sec.~\ref{subsec:streda_formula} shows the St\v{r}eda formulas for the SMOM.
In Sec.~\ref{sec:models}, we introduce three models for numerical calculations and in Sec.~\ref{sec:results}, we show the results with their interpretations.
Finally, Sec.~\ref{sec:conclusions} summarizes this work.

\section{Thermodynamic derivation of the SMOM} \label{sec:octupole_derivation}

\subsection{Thermodynamic definition of magnetic multipoles} \label{subsec:octupole_definition}

Our starting point is to construct a grand potential $\Omega[E,T,\mu] = E -T S(E,T,\mu) - \mu N(E,T,\mu)$ that includes a spin contribution.
Here, the thermodynamic functions $S$ and $N$ are the entropy and particle number, respectively, and their natural variables are the energy $E$, temperature $T$, and chemical potential $\mu$.
In general, spin is introduced into the Hamiltonian through its coupling with a Zeeman magnetic field $\bm{B}(\bm{r})$:
\begin{equation}
    \hat{H}_{\bm{B}} = -\frac{g_s e}{2m_e} \int d\bm{r} \hat{\bm{s}}(\bm{r}) \cdot \bm{B}(\bm{r}). \label{eq:zeeman}
\end{equation}
Here, $\hat{\bm{s}}(\bm{r}) = \frac{1}{2} \{ \hat{\bm{s}} , \hat{n}(\bm{r}) \}$ is the spin density operator, $\hat{\bm{s}}$ is the spin operator of the electron, and $\hat{n}(\bm{r}) = \delta(\bm{r} - \hat{\bm{r}})$ is the density operator.
In particular, $\hat{\bm{s}}$ is given by $\hbar \bm{\sigma}/ 2$, where $\hbar$ is the Dirac constant, and $\bm{\sigma}$ is the Pauli matrix vector.
The coefficient includes the electron charge $-e(<0)$, spin $g$-factor of the electron $g_s$ ($\approx 2$), and electron mass $m_e$, and is omitted for simplicity in the following.
Equation~\eqref{eq:zeeman} demonstrates that the effect of spin is incorporated through the energy $E$, and magnetic fields behave as natural variables conjugate to spin.

Next, we define magnetic multipoles using the grand potential.
For this purpose, we assume that the spatial modulation of the magnetic field $\bm{B}(\bm{r})$ is more gradual than the lattice constant. 
Performing the gradient expansion of this magnetic field, each gradient serves as a thermodynamic variable as
\begin{align}
    d \Omega =& -S dT - N d\mu - ( M_i dB_i + Q^{(\mathrm{m})}_{ij} d [\partial_j B_i] \nonumber \\ &+ O_{ijk} d[\partial_{jk} B_i]  + \cdots),\label{eq:thermodynamic_expansion}
\end{align}
where $\partial_j=\partial/\partial r_j$ and $\partial_{jk}=\partial^2/\partial r_j \partial r_k.$
The magnetic moments conjugate to these gradients denote magnetic multipoles.
Specifically, $M_i =  - \partial \Omega/\partial B_i$ is the spin magnetic dipole, $Q^{(\mathrm{m})}_{ij} = - \partial \Omega/\partial [\partial_j B_i]$ is the spin magnetic quadrupole, and $O_{ijk} = - \partial \Omega/\partial [\partial_{jk} B_i]$ is the spin magnetic octupole.

Then, Maxwell relations can be utilized for calculating these magnetic multipoles.
Specifically, for the spin magnetic octupole, the following Maxwell relation is established:
\begin{equation}
    \frac{\partial O_{ijk}}{\partial \mu} = \frac{\partial N}{\partial [\partial_{jk} B_i]}. \label{eq:maxwell}
\end{equation}
The usage of this relation has three key advantages.
The first advantage is that the right-hand side (r.h.s.) can be calculated by static linear response theory, which is more feasible than directly calculating the multipole through its thermodynamic definition.
After the calculation of the r.h.s., the multipole can be evaluated by integrating over the chemical potential.
The second advantage is that this equation avoids using the position operator \(\hat{\bm{r}}\), which is unbounded in bulk crystals and poses severe problems when straightforwardly calculating multipoles. 
The third advantage is that this equation is independent of the origin due to the absence of the position operator.

This thermodynamic approach has been developed within \textit{the modern theory of orbital magnetization} to define the orbital magnetic dipole moment~\cite{Shi2007-xs}, which faces similar challenges due to the position operator.
So far, it has also been applied to higher-rank multipoles, such as magnetic quadrupoles~\cite{Gao2018-uk,Shitade2019-hq,Shitade2018-rd,Gao2018-no} and electric quadrupoles~\cite{Daido2020-yo}, and has been established as a method for calculating multipoles in bulk crystals.

\subsection{Formulation of the SMOM} \label{subsec:octupole_expression}

In this subsection, we derive a reciprocal-space expression for the SMOM that can be evaluated by a Brillouin zone (BZ) integral of Bloch wave functions.
According to Eq.~\eqref{eq:maxwell}, the expression is obtained by expanding the density-spin correlation function $\Phi_i(\bm{q})$ up to the second order in momentum $\bm{q}$ and integrating it over the chemical potential $\mu$.
This correlation function is given by
\begin{align}
    \Phi_i(\bm{q}) &= \sum_{nm}\int_{\mathrm{BZ}} \frac{d^dk}{(2 \pi)^d}\frac{f_{n\bm{k}-\bm{q}/2} - f_{m\bm{k}+\bm{q}/2} }{ \epsilon_{n\bm{k}-\bm{q}/2} - \epsilon_{m\bm{k}+\bm{q}/2} } \notag \\
    & \quad \times \braket{u_{n\bm{k}-\bm{q}/2} | u_{m\bm{k}+\bm{q}/2} } 
    \bra{u_{m\bm{k}+\bm{q}/2}} \hat{s}^i \ket{u_{n\bm{k}-\bm{q}/2}},
\end{align}
where $d$ is the dimension, $\ket{u_{n\bm{k}}}$ is the periodic part of the Bloch wave function with momentum $\bm{k}$ and a band index $n$, $\epsilon_{n\bm{k}}$ is the band energy of the $n$th band, and $f_{n\bm{k}} = [1 + e^{(\epsilon_{n\bm{k}} - \mu)/T}]^{-1}$ is the Fermi distribution function.
Leaving the detailed derivation to Appendix~\ref{sec:derivation}, we present the final expression:
\begin{widetext}
\begin{align}
    O_{i,ab}
    =&
    \frac{1}{2}\sum_{n} \int_{\mathrm{BZ}} \frac{d^dk}{(2 \pi)^d}
    \mathrm{Re} \biggl[
    \sum_{m(\neq n)}
    \mathcal{G}_n \biggl(
    \frac{ (\mathrm{q}_{\mathrm{e}}^{ab})_{nm} (\mathrm{d}_{\mathrm{m}}^{i})_{mn} }{ \epsilon_{nm}}
    +
    \frac{2 (\mathrm{d}_{\mathrm{e}}^{a})_{nm} (\mathrm{q}_{\mathrm{m}}^{i,b})_{mn} }{ \epsilon_{nm}}
    \biggr)
    +\sum_{m(\neq n)}
    \biggl(
    \partial_b \mathcal{G}_n - \frac{\partial_b \tilde{\epsilon}_{nm} \mathcal{G}_n }{ \epsilon_{nm}}
    \biggr)\frac{i (\mathrm{d}_{\mathrm{e}}^{a})_{nm} (\mathrm{d}_{\mathrm{m}}^{i})_{mn}}{\epsilon_{nm}} \nonumber \\
    &\hspace{80pt}
    + \frac{f_n}{2} ( (\mathrm{o}_{\mathrm{m}}^{i,ab})_{n} + (\mathrm{q}_{\mathrm{e}}^{ab})_n (\mathrm{d}_{\mathrm{m}}^i)_n )
    +
    \frac{f'_n}{4} v^b_n \partial_a (\mathrm{d}_{\mathrm{m}}^i)_n
    +
    \frac{f''_n}{12} v^a_n v^b_n (\mathrm{d}_{\mathrm{m}}^i)_n
    \biggr] + (a \leftrightarrow b). \label{eq:octupole}
\end{align}
\end{widetext}
Note that we replace $O_{ijk}$ with $O_{i,ab}$ to distinguish the spin and spatial components and will often use this notation in the following.
Here, $a,b$ are also Cartesian components, $\partial_a=\partial/\partial k_a$ represents a momentum derivative, and $(a \leftrightarrow b)$ denotes the term that interchanges the spatial components in the first part.
Other notations appearing in Eq.~\eqref{eq:octupole} are as follows: $v^a_n = \partial_{a} \epsilon_{n\bm{k}} $ is the velocity operator, $\mathcal{G}_n = -\int^\mu_{-\infty} f_nd\mu' $ is the grand potential density resulting from the $\mu$-integration, and the following abbreviations are used:
$\epsilon_{nm} = \epsilon_{n\bm{k}} - \epsilon_{m\bm{k}}$ and $\tilde{\epsilon}_{nm} = \epsilon_{n\bm{k}} + \epsilon_{m\bm{k}}$.

Then, we confirm the gauge-invariance of Eq.~\eqref{eq:octupole} by analyzing the transition matrices,
\begin{subequations} \label{eq:transition}
    \begin{align}
        &(\mathrm{d}_{\mathrm{e}}^a)_{nm}
        =
        -i \braket{ u_{n\bm{k}} | D_a u_{m\bm{k}} }, \\
        &(\mathrm{d}_{\mathrm{m}}^i)_{nm} = \braket{ u_{n\bm{k}} | \hat{s}^i | u_{m\bm{k}}  }, \\
        &(\mathrm{q}_{\mathrm{e}}^{ab})_{nm} = \frac{1}{2} (
        \braket{ D_a u_{n\bm{k}}  | D_b u_{m\bm{k}}  }
        +
        \braket{ D_b u_{n\bm{k}}  | D_a u_{m\bm{k}}  }
        ), \\
        &(\mathrm{q}_{\mathrm{m}}^{i,a})_{nm}
        =
        \frac{-i}{2} (\braket{D_a u_{n\bm{k}} | \hat{s}^i | u_{m\bm{k}}} - \braket{ u_{n\bm{k}} | \hat{s}^i | D_a u_{m\bm{k}}}  ), \\
        &(\mathrm{o}_{\mathrm{m}}^{i,ab})_{n}
        =
        \frac{1}{2} (\braket{D_a u_{n\bm{k}} | \hat{s}^i | D_b u_{n\bm{k}} } +
        \braket{D_b u_{n\bm{k}} | \hat{s}^i | D_a u_{n\bm{k}} }).
\end{align}
\end{subequations}
Here, $D_a$ is the covariant derivative, which is defined as 
\begin{equation}
    \ket{D_a u_{n\bm{k}}} = \ket{\partial_a u_{n\bm{k}}} + iA^a_n \ket{u_{n\bm{k}}}, \label{eq:covariant_derivative}
\end{equation}
with the diagonal Berry connection $A^a_n = i \braket{u_{n\bm{k}} | \partial_a u_{n\bm{k}}}$.
This covariant derivative is gauge-covariant, which means that it transforms as $\ket{ D_a u_{n\bm{k}}} \to e^{i\beta_{n\bm{k}}} \ket{ D_a u_{n\bm{k}}}$ under a gauge transformation $\ket{u_{n\bm{k}}} \to e^{i\beta_{n\bm{k}}} \ket{u_{n\bm{k}}}$.
This property leads to the gauge-covariance of these transition rate matrices and finally to the gauge-invariance of Eq.~\eqref{eq:octupole}.

\subsection{Molecular limit} \label{subsec:molecular_limit}

In this subsection, we convert Eq.~\eqref{eq:octupole} to a real-space expression to clarify its physical meaning. 
Specifically, we replace the Bloch orbitals with molecular orbitals and take the molecular limit, where the electron wave function is well-localized in a unit cell.

For an ideal crystal, where wave functions do not overlap among different unit cells, the cell-periodic Bloch wave function can be written as~\cite{Pozo_Ocana2023-fk}
\begin{equation}
    u_{n\bm{k}}(\bm{r}) \approx e^{- i \bm{k} \cdot \bm{\xi}(\bm{r}) } \phi_n(\bm{\xi}(\bm{r})). \label{eq:wave_function}
\end{equation}
Here, $\phi_n(\bm{r})$ are the molecular orbitals in a unit cell, and $n$ is an index of the eigenstates.
If the atoms in the cell are sufficiently spaced apart, the index $n$ designates the atomic orbitals at each atomic site. 
The vector $\bm{\xi}(\bm{r}) = \bm{r} - \bm{R}(\bm{r})$ is the intracell coordinate, and $\bm{R}(\bm{r})$ is a lattice vector that maps the absolute coordinate $\bm{r}$ back to the home unit cell.
The exact equality of Eq.~\eqref{eq:wave_function} is established in the molecular limit, where the diagonal Berry connection becomes
\begin{equation}
    A_n^a = i \int_{\mathrm{cell}} d\bm{r} u^*_{n\bm{k}}(\bm{r}) \partial_{a} u_{n\bm{k}}(\bm{r})
    =
    \braket{ \phi_n | \hat{r}^a | \phi_n } =: \bar{r}_n.
\end{equation}
Furthermore, the covariant derivative is 
\begin{equation}
    D_a u_{n\bm{k}}(\bm{r})
    =
    -i e^{-i \bm{k} \cdot \bm{r}} (r^a - \bar{r}^a_n) \phi_n(\bm{r}),
\end{equation}
where $\bm{r}$ is in the home unit cell.

With these equations, the transition rate matrices in Eqs.~\eqref{eq:transition} are expressed as
\begin{subequations}
    \begin{align}
        &(\mathrm{d}_{\mathrm{e}}^a)_{nm}
        =
        \braket{ \phi_n | \hat{r}^a - \bar{r}^a_{n+m}/2  | \phi_m }, \\
        &(\mathrm{d}_{\mathrm{m}}^i)_{nm} = \braket{ \phi_n | \hat{s}^i | \phi_m  }, \\
        &(\mathrm{q}_{\mathrm{e}}^{ab})_{nm} = 
        \braket{ \phi_n | (\hat{r}^a - \bar{r}^a_{n+m}/2)(\hat{r}^b - \bar{r}^b_{n+m}/2)  | \phi_m }, \\
        &(\mathrm{q}_{\mathrm{m}}^{i,a})_{nm}
        =
        \braket{ \phi_n | (\hat{r}^a - \bar{r}^a_{n+m}/2) \hat{s}^i  | \phi_m }
        , \\
        &(\mathrm{o}_{\mathrm{m}}^{i,ab})_{n}
        =
       \braket{ \phi_n | (\hat{r}^a - \bar{r}^a_{n})(\hat{r}^b - \bar{r}^b_{n}) \hat{s}^i  | \phi_n },
    \end{align}    
\end{subequations}
where we define $\bar{r}^a_{n+m} = \bar{r}^a_{n} + \bar{r}^a_{m}$.
These matrices are independent of the origin of the coordinate system, which is ensured by the covariant derivative.
Furthermore, they represent the transition rates of multipoles, listed in order from top to bottom as electric dipole (E1), magnetic dipole (M1), electric quadrupole (E2), magnetic quadrupole (M2), and magnetic octupole (M3).
Therefore, Eq.~\eqref{eq:octupole} consists of the following components: combinations of the transition rate matrices, E2-M1 (first term) and E1-M2 (second term), the expectation value of M3 (fifth term), and a composite order formed by E2 and M1 (sixth term).
In particular, the off-diagonal components of E2-M1 and E1-M2 are finite only in multi-atomic orbital systems, leaving only the fifth and sixth terms in single-atomic orbital systems.
The remaining terms in Eq.~\eqref{eq:octupole} can be regarded as unique contributions to crystals, considering that the group velocity arises from $\partial_b\mathcal{G}_n$ (third term) and $\partial_b\tilde{\epsilon}_{nm}$ (fourth term).
This is because the band dispersion disappears in the molecular limit, resulting in a zero group velocity.

\subsection{St\v{r}eda formulas} \label{subsec:streda_formula}

St\v{r}eda formulas describe the relationship between a multipole and its corresponding response coefficient in insulators at zero temperature.
The original St\v{r}eda formula, proposed by P. St\v{r}eda, relates orbital magnetization to the Hall coefficient~\cite{Streda1982-xz}.
In this subsection, we establish the St\v{r}eda formulas between the SMOM and four response coefficients.

The first two equations can be derived by analyzing the following response phenomena,
\begin{align}
    P_i=\chi^{(\mathrm{d_e q_m})}_{ijk}\partial_kB_j, \quad Q^{(\mathrm{e})}_{ij}=\chi^{(\mathrm{q_e d_m})}_{ijk}B_k, \label{eq:response1}
\end{align}
where $P_i$ is an electric polarization, $Q^{(\mathrm{e})}_{ij}$ is an electric quadrupole, and $B_i$ is a magnetic field.
The left and right equations express the electric polarization response induced by a magnetic field gradient and the electric quadrupole response induced by a magnetic field, respectively.
The corresponding response coefficients $\chi_{ijk}$ are symbolically represented by $(\mathrm{d_e q_m})$ and $(\mathrm{q_e d_m})$, respectively.
These induced electric multipoles contribute to the charge density in insulators at zero temperature as $\rho=eN=-\partial_i(P_i-\partial_jQ^{(\mathrm{e})}_{ij})$. 
This relation can be derived from the multipole expansion of the scalar potential $\phi$ in an insulating medium, with the quadrupole term emerging upon including higher-order terms~\cite{book1}.
Substituting Eq.~\eqref{eq:response1} into this expression and using Eq.~\eqref{eq:maxwell}, we obtain the following two St\v{r}eda formulas:
\begin{align}
    e\frac{\partial O_{ijk}}{\partial \mu}=-\chi^{(\mathrm{d_e q_m})}_{kij}, \quad e\frac{\partial O_{ijk}}{\partial \mu}=\chi^{(\mathrm{q_e d_m})}_{jki}. \label{eq:streda1}
\end{align}

Similarly, two more equations can be derived by analyzing the following response phenomena,
\begin{align}
    M_i=\chi^{(\mathrm{d_m q_e})}_{ijk}\partial_kE_j, \quad Q^{(\mathrm{m})}_{ij}=\chi^{(\mathrm{q_m d_e})}_{ijk}E_k, \label{eq:response2}
\end{align}
where $E_i$ is an electric field.
The left and right expressions describe the spin magnetization response induced by an electric field gradient and the spin magnetic quadrupole response induced by an electric field, respectively.
Their response coefficients $\chi_{ijk}$ are symbolically denoted as $(\mathrm{d_m q_e})$ and $(\mathrm{q_m d_e})$, respectively.
According to Eq.~\eqref{eq:thermodynamic_expansion}, these induced magnetic multipoles are expressed as
\begin{align}
    M_i&=\partial_j\partial_kO_{ijk}=\partial_k\biggl[\partial_j\mu \frac{\partial O_{ijk}}{\partial \mu}\biggr], \\
    Q^{(\mathrm{m})}_{ij} &=-\partial_kO_{ijk} =-\partial_k\mu \frac{\partial O_{ijk}}{\partial \mu},
\end{align}
where we only focus on the SMOM component.
Assuming an insulator at zero temperature and using the relation $\partial_j\mu=-e\partial_j\phi=eE_j$, we obtain the following two St\v{r}eda fomulas:
\begin{align}
    e\frac{\partial O_{ijk}}{\partial \mu}=\chi^{(\mathrm{d_m q_e})}_{ijk}, \quad e\frac{\partial O_{ijk}}{\partial \mu}=-\chi^{(\mathrm{q_m d_e})}_{ijk}. \label{eq:streda2}
\end{align}

These four response coefficients are related to each other through Maxwell relations:
\begin{align}
    \chi^{(\mathrm{d_e q_m})}_{kij}=\frac{\partial P_k}{\partial [\partial_jB_i]}&=\frac{\partial Q^{(\mathrm{m})}_{ij}}{\partial E_k}=\chi^{(\mathrm{q_m d_e})}_{ijk}, \\
    \chi^{(\mathrm{q_e d_m})}_{jki}=\frac{\partial Q^{(\mathrm{e})}_{jk}}{\partial B_i}&=\frac{\partial M_i}{\partial [\partial_kE_j]}=\chi^{(\mathrm{d_m q_e})}_{ijk}.
\end{align}
Therefore, we uniformly refer to these coefficients as \textit{the spin magnetoelectric dipole-quadrupole susceptibilities.}
In summary, the SMOM is related to these susceptibilities through the St\v{r}eda formulas. 
This correspondence is akin to the relationship between the magnetic quadrupole moment and magnetoelectric dipole-dipole susceptibility  (commonly referred to as the magnetoelectric susceptibility)~\cite{Gao2018-uk,Shitade2019-hq,Shitade2018-rd,Gao2018-no}.
Note that the above discussion can also be interpreted as the St\v{r}eda formulas for the orbital MOM.
In this context, the orbital MOM corresponds to the orbital magnetoelectric dipole-quadrupole susceptibilities.

\section{Models} \label{sec:models}

In this section, we introduce three models to calculate the SMOM.
The first model is a pyrochlore lattice with an all-in/all-out magnetic configuration.
This model has four sublattice degrees of freedom and exhibits a cluster-scale magnetic octupole that spans across the sublattices~\cite{Arima2013-mp}.
The Hamiltonian is given by~\cite{Hayami2019-zk,Yamaji2014-fh}
\begin{align}
    H_{\mu \nu}(\bm{k})&=2 \bigl[-t+i\lambda(\bm{d}_{\mu \nu}\cdot \bm{\sigma})\bigr] \cos(\bm{b}_{\mu \nu}\cdot \bm{k})+m \bm{Q}_{\mu\nu}\cdot \bm{\sigma},
\end{align}
where $\mu,\nu$ are the indices of the sublattices, and $\bm{\sigma}=(\sigma_x,\sigma_y,\sigma_z)$ are the Pauli matrices of the spin degrees of freedom.
The coefficients $t$ and $\lambda$ denote the hopping and SOC between nearest-neighbor sites.
The vectors are given by $\bm{d}_{\mu \nu}=2\bm{a}_{\mu \nu}\times\bm{b}_{\mu \nu}$, $\bm{a}_{\mu \nu}=(\bm{x}_\mu+\bm{x}_\nu)/2-\bm{x}_{\mathrm{G}}$, and $\bm{b}_{\mu\nu}=\bm{x}_\nu-\bm{x}_\mu$, where $\bm{x}_\mu$ is the position vector at a sublattice $\mu$ in a unit tetrahedron, and $\bm{x}_{\mathrm{G}}$ is the center of the tetrahedron.  
The diagonal matrices $\bm{Q}=(Q^x,Q^y,Q^z)$ denote on-site charge degrees of freedom, and the product of $\bm{Q}$ and $\bm{\sigma}$ represents a magnetic octupole.
In particular, the all-in/all-out configuration requires $Q^x=\mathrm{diag}(1,-1,-1,1),Q^y=\mathrm{diag}(1,-1,1,-1)$, and $Q^z=\mathrm{diag}(1,1,-1,-1)$, and $m$ reflects the magnitude of the order parameter.

The second and third models are a two-dimensional (2D) altermagnet with $d_{x^2-y^2}$ symmetry and a three-dimensional (3D) altermagnet with $d_{xy}$ symmetry, which exhibit an atomic-scale magnetic octupole~\cite{Bhowal2024-vj}.
The Hamiltonians are given by~\cite{Roig2024-zg,Antonenko2025-jp}
\begin{align}
    H(\bm{k})=\epsilon_0+t_x\tau_x+t_z\tau_z+\tau_y \bm{\lambda}\cdot\bm{\sigma}+\tau_z\bm{J}\cdot\bm{\sigma},
\end{align}
where $\bm{\tau}=(\tau_x,\tau_y,\tau_z)$ and $\bm{\sigma}=(\sigma_x,\sigma_y,\sigma_z)$ are the Pauli matrices of the sublattice and spin degrees of freedom, respectively.
The coefficient $\epsilon_0$ describes a sublattice-independent dispersion, and $t_x$ and $t_z$ denote the inter- and intra-sublattice hoppings, respectively.
The vectors $\bm{\lambda}=\lambda\bm{g}$ and $\bm{J}=J\bm{N}$ represent the SOC and a staggered effective field, respectively.
Here, $\lambda$ and $J$ are the coupling constants, $\bm{g}$ is the direction vector of the SOC, and $\bm{N}$ is the N\'{e}el vector.
The detailed expressions of each term are given in Appendix~\ref{subsec:model_detail}.
In the following, we assume $\bm{g}=(0,0,g_z)$ and $\bm{N}=(0,0,N)$ for the 2D altermagnet model and $\bm{g}=(g_x,g_y,0)$ and $\bm{N}=(0,0,1)$ for the 3D altermagnet model.
In particular, the direction of $\bm{N}$ in the 3D model corresponds to that of an altermagnetic candidate RuO$_2$~\cite{Berlijn2017-yg,Zhu2019-fx}.

\begin{figure}[t]
    \setcounter{figure}{1} 
    \centering
    \subfigure[$N/N_c=0.5$]{
        \includegraphics[height=2.4cm,width=0.34\hsize]{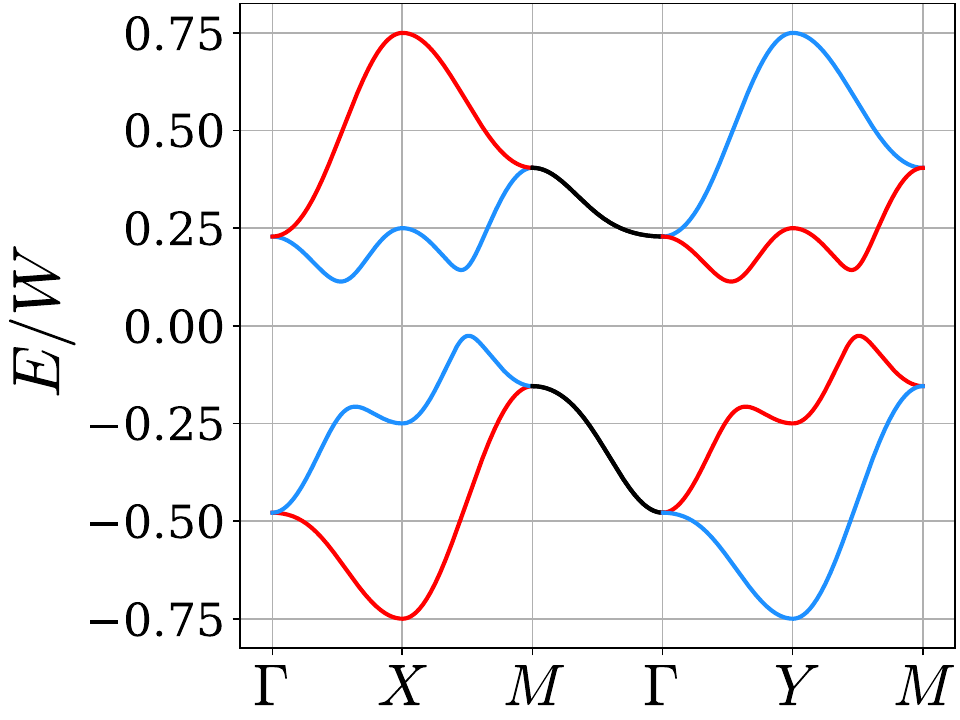}
        \label{fig:Altermagnet_Energy1}
    }
    \hfill
    \subfigure[$N/N_c=1.0$]{
        \includegraphics[height=2.4cm,width=0.28\hsize]{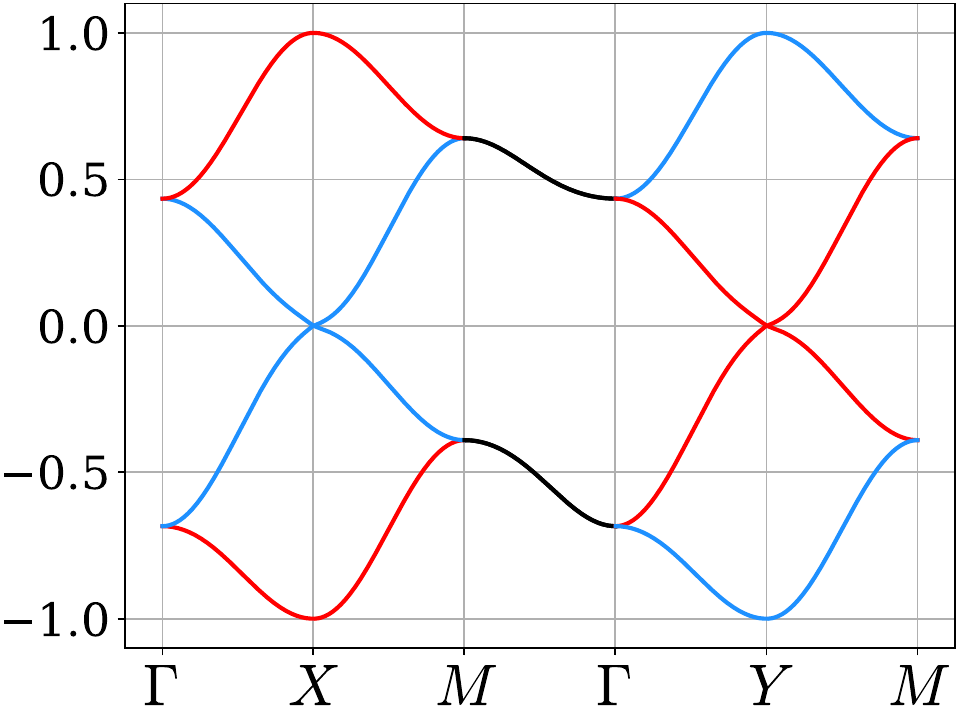}
        \label{fig:Altermagnet_Energy2}
    }
    \hfill
    \subfigure[$N/N_c=1.25$]{
        \includegraphics[height=2.4cm,width=0.28\hsize]{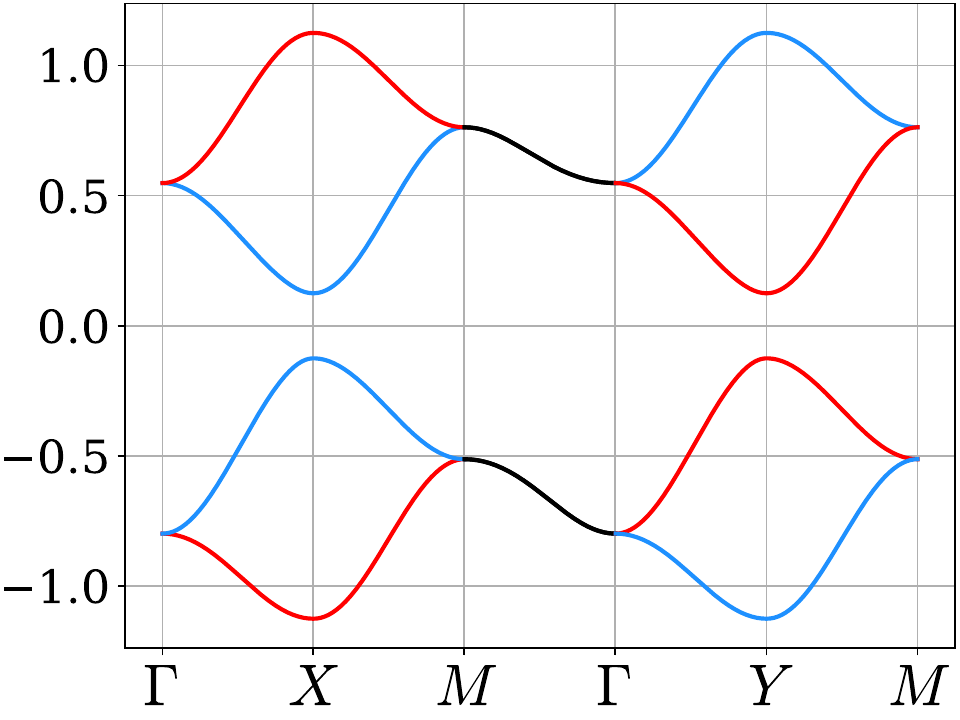}
        \label{fig:Altermagnet_Energy3}
    }
    \setcounter{figure}{0}
    \caption{Band dispersions of the 2D altermagnet model along high-symmetry lines for (a) $N/N_c=0.5$, (b) $N/N_c=1.0$, and (c) $N/N_c=1.25$.
    The red and blue lines represent the up-spin and down-spin bands, respectively.
    The region of $N/N_c<1$ is a nontrivial regime, and $N/N_c>1$ is a trivial regime, where the $X$ and $Y$ points are gap-closing points.
    Furthermore, this model exhibits spin degeneracy along the $k_x=\pm k_y$ lines (e.g., the $\Gamma$-$M$ line), which reflects $d_{x^2-y^2}$ symmetry.
    Here, we use the same parameters as in the numerical calculations and set $W=8t_d$.
    }
    \label{fig:Altermagnet}
\end{figure}

Notably, the 2D altermagnet model has a nontrivial band topology, and the parameter $N$ determines its presence or absence~\cite{Antonenko2025-jp}. 
The energy dispersion is given by [Appendix~\ref{subsec:model_detail}]
\begin{align}
    E^{\uparrow \downarrow}_{\pm}=\epsilon_0 \pm \sqrt{t_x^2+(t_z+\sigma JN)^2+(\lambda g_z)^2}, \label{eq:Altermagnet_models_energy2D}
\end{align}
where $\sigma=+1$ for the up-spin ($\uparrow$) band and $\sigma=-1$ for the down-spin ($\downarrow$) band.
In particular, the energies at the $X(\pi,0)$ and $Y(0,\pi)$ points ar
\begin{align}
    E^{\downarrow}_{\pm}(X)=\pm J|N-N_c|, \quad
    E^{\uparrow}_{\pm}(Y)=\pm J|N-N_c|,  \label{eq:Altermagnet_models_energy2D_XY}
\end{align}
where $N_c=4t_d/J$, and $t_d$ depends on $t_z$.
These equations demonstrate that the bands become inverted near the $X$ and $Y$ points depending on $N$, as shown in Figs.~\ref{fig:Altermagnet_Energy1}-\ref{fig:Altermagnet_Energy3}.
Consequently, this model exhibits a topological phase transition at $N=N_c$.

Next, we consider the symmetry constraints of the SMOM in these models.
The details can be found in Appendix~\ref{subsec:symmetry_detail}, and brief summaries are given below.
The SMOM, which is a third-rank time-reversal-odd axial tensor, obeys the following transformation rule~\cite{Xiao2022-xr}:
\begin{align}
    O_{i',a'b'}=\eta_T\mathrm{det}(\mathcal{R})\mathcal{R}_{i'i}\mathcal{R}_{a'a}\mathcal{R}_{b'b}O_{i,ab}, \label{eq:classification}
\end{align}
where $\mathcal{R}$ is a symmetry operation, and $\eta_T=\pm 1$ for unitary ($+$) and anti-unitary ($-$) operations.
The magnetic point group of the pyrochlore model is $m\bar{3}m'$, which includes a $2\pi/3$ rotation around the [1,1,1] axis $C^{[111]}_3: (x,y,z) \rightarrow(z,x,y)$ and two mirror symmetries $\mathcal{M}_{(100)}: x \rightarrow -x$ and $\mathcal{M}_{(001)}: z \rightarrow -z$.
The $C^{[111]}_3$ symmetry enforces $O_{x,yz}=O_{y,zx}=O_{z,xy}$, as well as additional equivalent relations. 
However, $\mathcal{M}_{(100)}$ and $\mathcal{M}_{(001)}$ only preserve $O_{x,yz}=O_{y,zx}=O_{z,xy}$.
Therefore, we calculate the only independent component denoted as $O_{x,yz}$ in the pyrochlore model.

On the other hand, the magnetic point group of the altermagnet models is $4'/mm'm$, which is characterized by $C^{[001]}_4\mathcal{T}$, a combined operation of a $\pi/2$ rotation around the $z=[0,0,1]$ axis $C^{[001]}_4: (x,y,z) \rightarrow (-y,x,z)$ and time reversal $\mathcal{T}$.
This $C^{[001]}_4 \mathcal{T}$ symmetry imposes the following relations: $O_{x,yz}=O_{y,zx}$, $O_{x,zx}=-O_{y,yz}$, and $O_{z,xx}=-O_{z,yy}$. 
Additionally, it leaves another independent component $O_{z,xy}$.
Together with this operation, the mirror symmetries in each model determine the final symmetry constraints of the SMOM.
Specifically, $\mathcal{M}_{(110)}: (x,y)\rightarrow(-y,-x)$ in the 2D model enforces $O_{z,xx}=-O_{z,yy}$, and $\mathcal{M}_{(100)}$ and $\mathcal{M}_{(010)}: y \rightarrow -y$ in the 3D model lead to $O_{x,yz}=O_{y,zx}\neq O_{z,xy}$.
Therefore, we calculate $O_{z,xx}$ in the 2D model and $O_{x,yz}$ and $O_{z,xy}$ in the 3D model.

\section{Numerical results} \label{sec:results}

In this section, we calculate the SMOM for the three models introduced in the previous section and show the results in Secs.~\ref{subsec:Pyrochlore}, \ref{subsec:2DAltermagnet}, and \ref{subsec:3DAltermagnet}, respectively.
Unless otherwise noted, the temperature is set to $T=0.01$ for the pyrochlore and 3D altermagnet models and to $T=0.001$ for the 2D altermagnet model.
Note that all computed SMOMs are expressed in units of $\mu_{\mathrm{B}}\mathrm{\AA}^2$, where $\mu_{\mathrm{B}}=e\hbar/2m_e$ is the Bohr magneton.

Before showing the calculation results, we note the following two important points: 
First, Eq.~\eqref{eq:octupole} is unsuitable for numerical calculations because it includes derivatives of the Bloch wave function.
Therefore, we apply the Hellmann-Feynman theorem $\braket{u_{n\bm{k}}|D_a u_{m\bm{k}}}=-v^a_{nm}/\epsilon_{nm}$ and rewrite it with the velocity operator $v^a_{nm}=\braket{u_{n\bm{k}}|\partial_aH_{\bm{k}}|u_{m\bm{k}}}$ and spin operator $s^i_{nm}=(\mathrm{d}_{\mathrm{m}}^i)_{nm}$.
The rewritten expression is given in Appendix~\ref{sec:SMOM_numerical}.

Second, we discuss how to handle degenerate points when calculating the SMOM.
The expression includes an infinitesimal quantity $\Delta \epsilon\ (\ll 1) $ such as $\epsilon_{nm}$ in the denominator and seems to diverge at degenerate points.
However, this divergence does not actually occur because $O_{i,ab}$ is a thermodynamic quantity.
Indeed, the absence of a divergence can be confirmed by expanding the expression in terms of $\Delta\epsilon$, which leads to the cancellation of the $\Delta \epsilon^{-k}(k>1)$ terms [see Appendix~\ref{sec:SMOM_degeneracy}].
The remaining terms, for which convergence is guaranteed, can be used in numerical calculations around degenerate points.
Here, we adopt the expanded expression up to the first order of $\Delta \epsilon$ ($\Delta \epsilon^{0}$ and $\Delta \epsilon^{1}$ terms).
Note that we apply this expression only to the 2D and 3D altermagnet models with spin degeneracy [see Figs.~\ref{fig:Altermagnet_Energy1}-\ref{fig:Altermagnet_Energy3}] and not to the pyrochlore model.
As a criterion for whether to apply this expression, we set $\Delta\epsilon_c=10^{-4}$ for the 2D model and $\Delta\epsilon_c=10^{-6}$ for the 3D model.

\begin{figure}[t]
    \setcounter{figure}{2} 
    \centering
    \subfigure[]{
        \includegraphics[height=3.0cm,width=0.46\hsize]{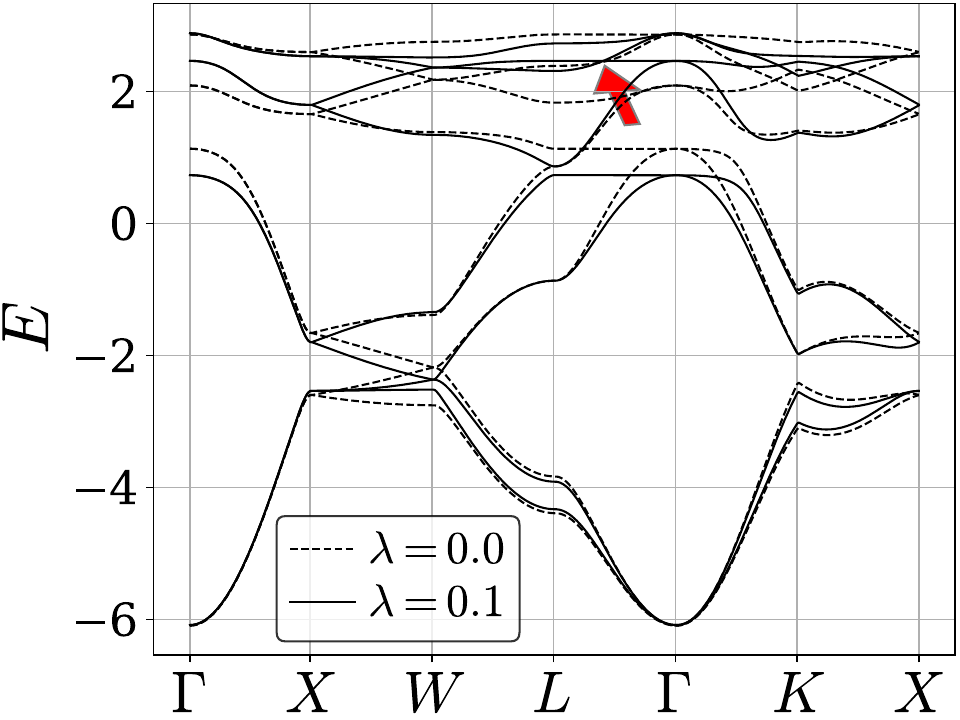}
        \label{fig:Pyrochlore_Energy_metal}
    }
    \hfill
    \subfigure[]{
        \includegraphics[height=3.2cm,width=0.46\hsize]{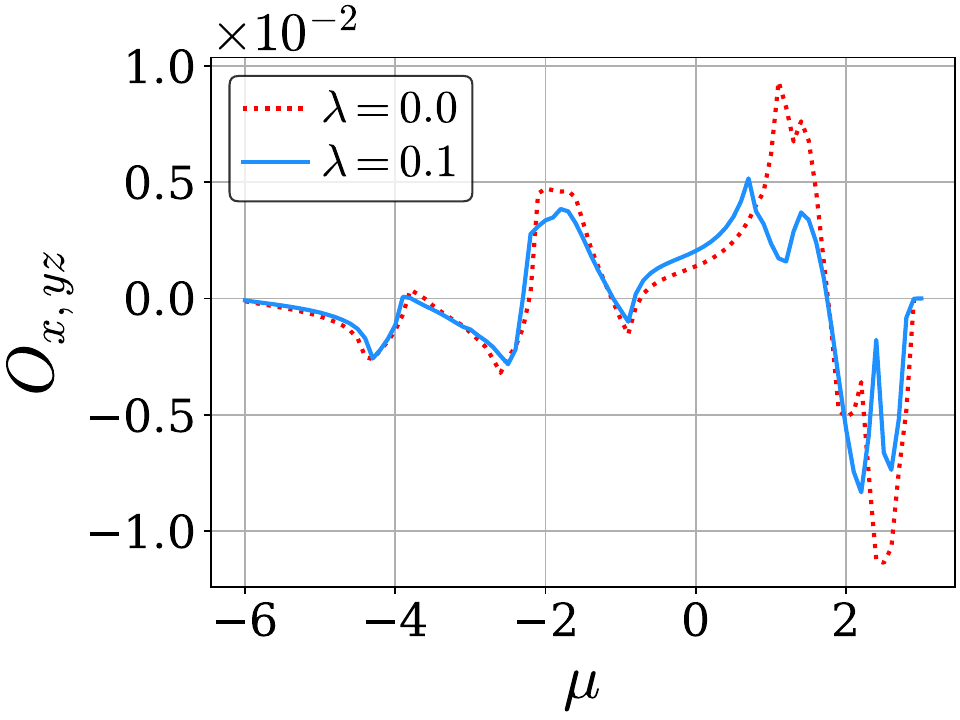}
        \label{fig:Pyrochlore_Octupole_metal}
    }
    \subfigure[]{
        \includegraphics[height=3.0cm,width=0.46\hsize]{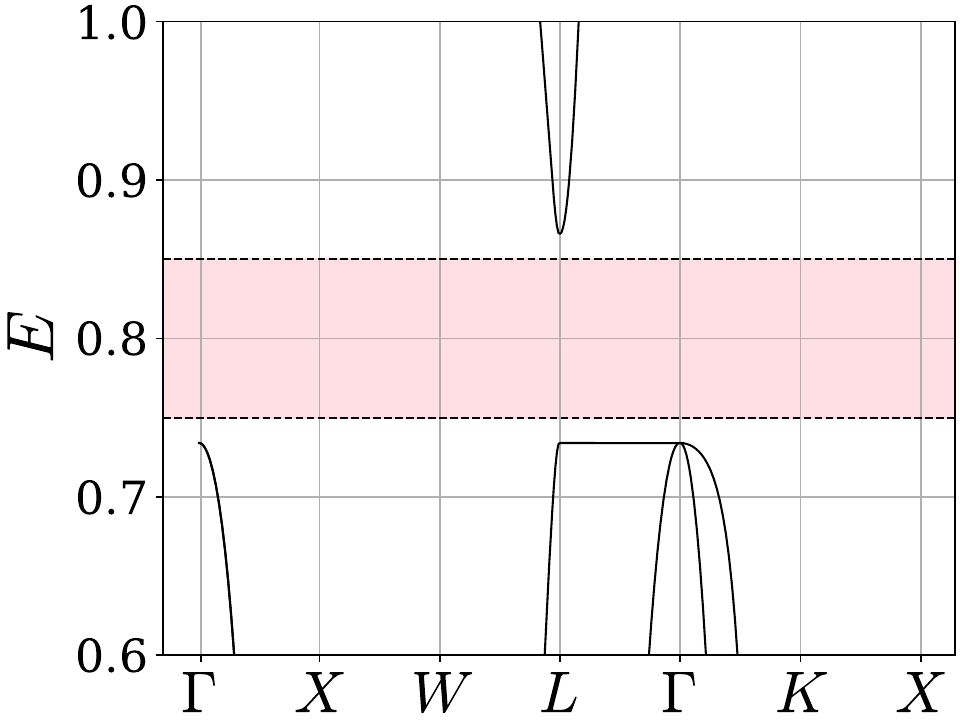}
        \label{fig:Pyrochlore_Energy_insulator}
    } 
    \hfill
    \subfigure[]{
        \includegraphics[height=3.2cm,width=0.46\hsize]{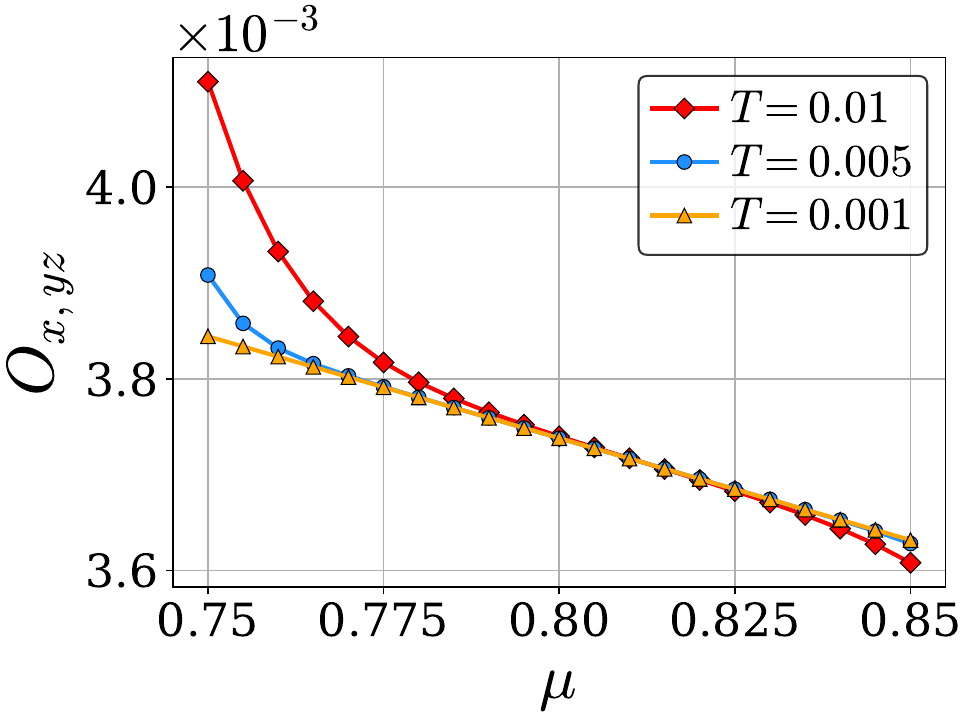}
        \label{fig:Pyrochlore_Octupole_insulator}
    } 
    \setcounter{figure}{1}
    \caption{ 
            (a) Band dispersion of the pyrochlore model for $(t,m)=(1.0,0.5)$ without SOC (dashed line) and with SOC (solid line).
            This model has Weyl points around $E=2.5$ in the presence of SOC, as indicated by the red arrow.
            (b) Chemical potential dependence of the SMOM without the SOC (red line) and with the SOC (blue line). 
            A noticeable enhancement does not appear around $\mu=2.5$. 
            (c) Magnified view of (a) with SOC. 
                An insulating phase exists in the region of $0.75 \le E \le 0.85$ (red-shaded area).
            (d) Chemical potential dependence of the SMOM in the insulating region at several temperatures $T$.
             In the numerical calculations, the other parameters are set to $\lambda=0.1$ in (d) and $(t,m)=(1.0,0.5)$ in both (b) and (d).}
    \label{fig:Pyrochlore}
\end{figure}

\subsection{Pyrochlore model} \label{subsec:Pyrochlore}

One aspect potentially relevant to the SMOM is quantum geometry. 
Our recent publication revealed that the intrinsic nonlinear magnetoelectric effect (INMEE), which is characterized by magnetic octupole order, is driven by the quantum metric and is enhanced near Weyl points~\cite{Oike2024-bl}. 
The SMOM also includes terms that involve the quantum metric (e.g., term (V) in Appendix~\ref{sec:SMOM_numerical}). 
This leads to an important question: Is the SMOM also enhanced at Weyl points, or is the enhancement of the INMEE at Weyl points independent of the SMOM?

The answer is that the SMOM does not necessarily exhibit an enhancement, unlike the INMEE.
We demonstrate this fact by examining the effect of Weyl points on the SMOM in the pyrochlore model [Fig.~\ref{fig:Pyrochlore_Energy_metal}].
This model features Weyl points in the presence of SOC~\cite{Oike2024-bl}, prompting us to investigate how the SMOM differs with or without SOC.
Specifically, we calculate the chemical potential dependence of the SMOM in the presence and absence of the SOC, as shown in Fig.~\ref{fig:Pyrochlore_Octupole_metal}.
However, the SMOM exhibits no significant differences and is actually larger in the absence of the SOC. 
This may be because the SMOM cannot be expressed solely in metric terms, unlike the INMEE tensor.

Furthermore, we investigate the St\v{r}eda formulas for the SMOM in an insulating region [Fig.~\ref{fig:Pyrochlore_Energy_insulator}] at several temperatures.
Figure~\ref{fig:Pyrochlore_Octupole_insulator} shows the chemical potential dependence of the SMOM in this region.
The SMOM exhibits a strong dependence on $\mu$ at high temperatures, whereas it follows a linear relationship at low temperatures, thereby confirming the validity of the Středa formulas.
Moreover, the slope of the line enables the estimation of the magnitudes of the corresponding response coefficients.

\begin{figure}[b]
    \setcounter{figure}{3} 
    \centering
    \subfigure[]{
        \includegraphics[height=3.2cm,width=0.51\hsize]{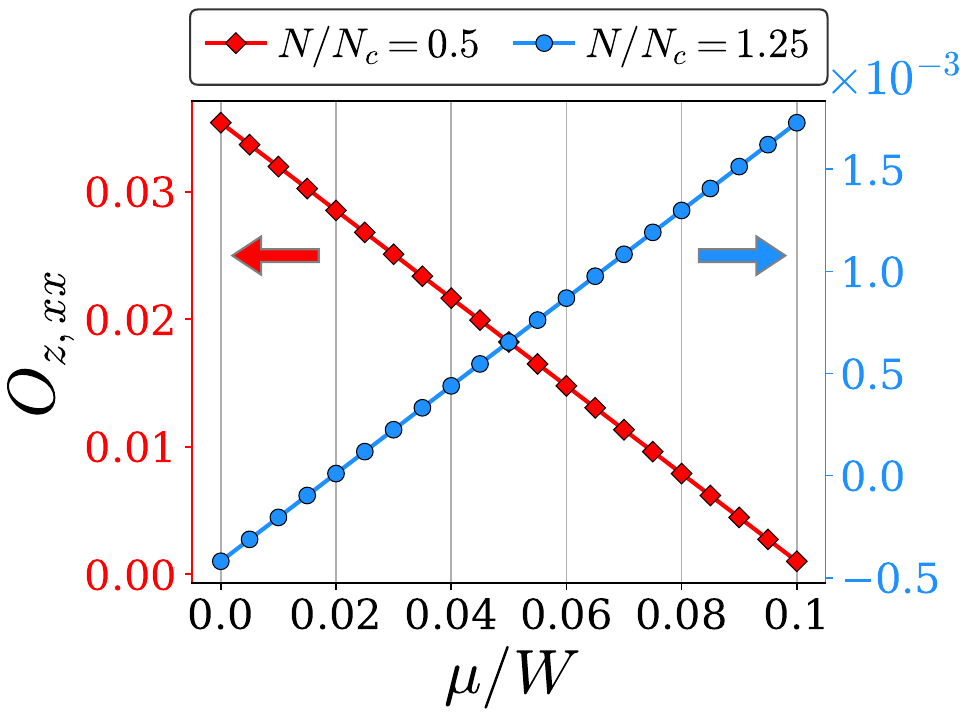}
        \label{fig:Altermagnet_Octupole_insulator}
    } 
    \hfill
    \subfigure[]{
        \includegraphics[height=3.2cm,width=0.428\hsize]{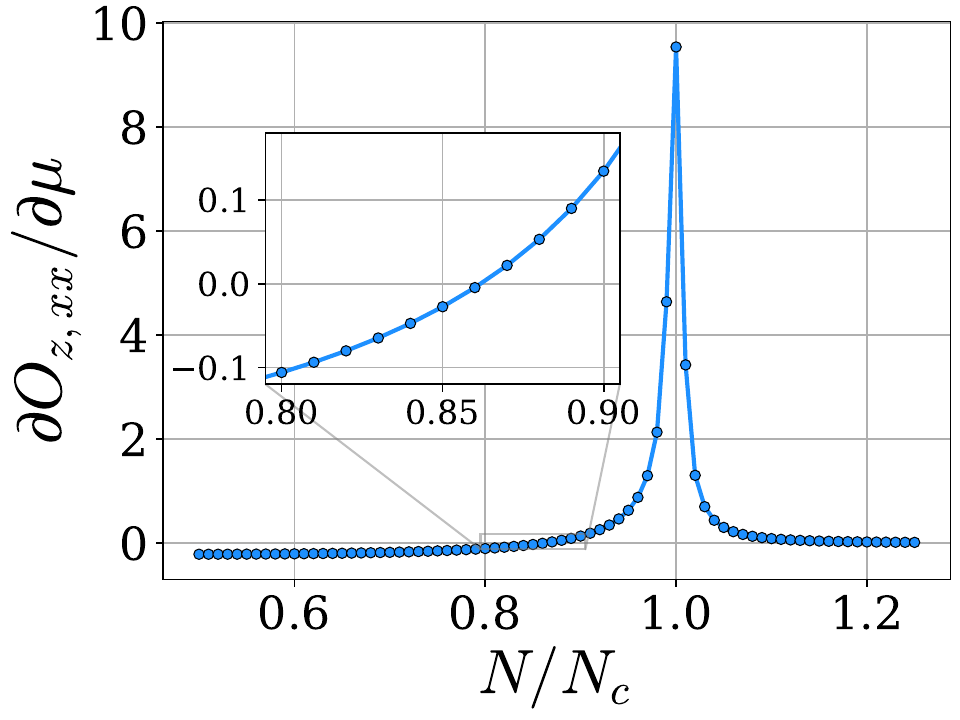}
        \label{fig:Altermagnet_Octupole_slope}
    } 
    \setcounter{figure}{2}
    \caption{(a) Chemical potential dependence of the SMOM in the 2D altermagnet model for $N/N_c=0.5$ (red line) and $N/N_c=1.25$ (blue line).
             (b) $N$ dependence of the slope ($\partial O_{z,xx}/\partial \mu$). 
             The inset is a magnified view of the region highlighted in the main panel.
             The sign of the slope changes within this range.}
        \label{fig:Altermagnet_2D}
\end{figure}

\subsection{2D altermagnet model} \label{subsec:2DAltermagnet}

The original St\v{r}eda formula connects topology and orbital magnetization in 2D insulating systems at zero temperature~\cite{Thouless1982-zh,Kohmoto1985-zj}, suggesting that topology might also affect the St\v{r}eda formulas for the SMOM.
Thus, we examine the St\v{r}eda formulas in nontrivial ($N/N_c<1$) and trivial ($N/N_c>1$) insulating regimes of the 2D altermagnet model.
For this purpose, guided by the result in Fig.~\ref{fig:Pyrochlore_Octupole_insulator}, we set the temperature to a low value of $T=0.001$.
Figure~\ref{fig:Altermagnet_Octupole_insulator} shows the chemical potential dependence of the SMOM in nontrivial and trivial regimes.
In both regimes, the SMOM exhibits a linear relationship except for the sign change of the slope.
To investigate this sign change, we further calculate the magnitude of the slope ($\partial O_{z,xx}/\partial \mu$) for various values of $N$, as shown in Fig.~\ref{fig:Altermagnet_Octupole_slope}.
The sign of the slope changes not exactly at the topological transition but rather before and after it [see the inset].
Therefore, this sign reversal may not be indicative of the transition.
Meanwhile, it shows a strong enhancement near the transition.

This enhancement is due to interband effects.
To confirm this result, we calculate the chemical potential dependence of each term in Eq.~\eqref{eq:octupole}, yielding the SMOM, as shown in Fig.~\ref{fig:Altermagnet_terms}. 
The main contributions come from the Fermi sea terms ($\propto f_n$) and a two-band term proportional to the grand potential density ($\propto \mathcal{G}_n$),
\begin{align}
    -\sum_n \int_{\mathrm{BZ}} \frac{d^dk}{(2\pi)^d} \sum_{m(\neq n)}
    \mathcal{G}_{n}\mathrm{Re} \biggl[ \frac{v^x_{nm}v^x_{mn}}{\epsilon^3_{nm}} \biggl](s^z_n+s^z_m), 
    \label{eq:interband}
\end{align}
which determines the slope.
Equation~\eqref{eq:interband} demonstrates that a small band gap of $\Delta\epsilon$ can enhance the slope by a factor of $1/\Delta \epsilon^3$.
Indeed, the momentum-resolved quantity of $\partial O_{z,xx}/\partial\mu$ becomes larger around points with small band gaps, such as the $X$ and $Y$ points, as shown in Fig.~\ref{fig:Altermagnet_color}.

\begin{figure}[t]
    \setcounter{figure}{4} 
    \centering
    \subfigure[]{
        \includegraphics[height=3.2cm,width=0.44\hsize]{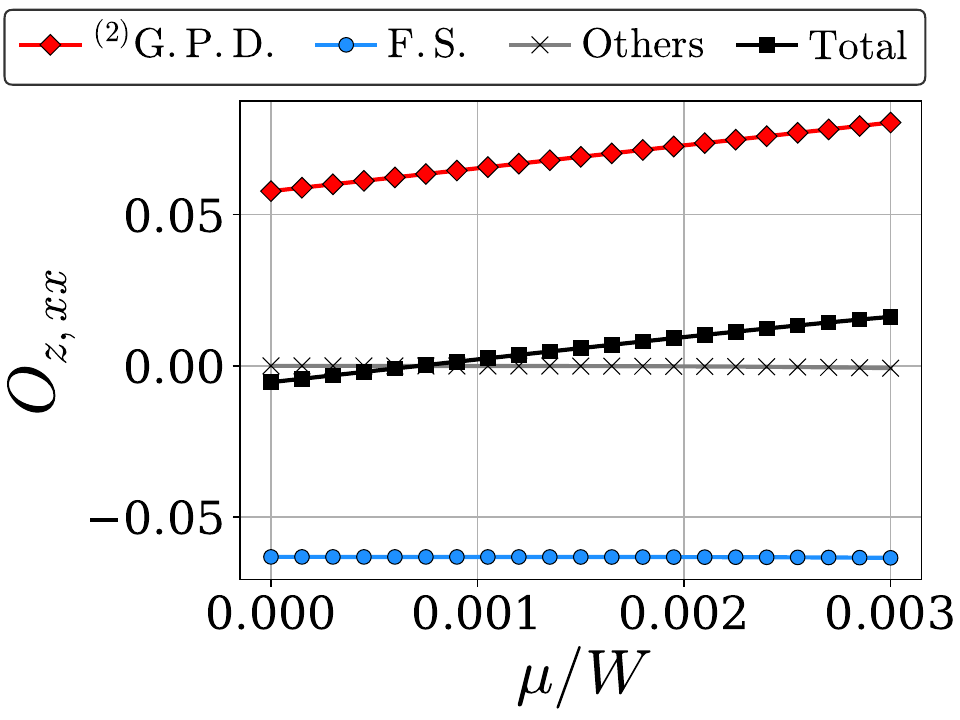}
        \label{fig:Altermagnet_terms}
    }
    \hfill
    \subfigure[]{
        \includegraphics[height=3.2cm,width=0.48\hsize]{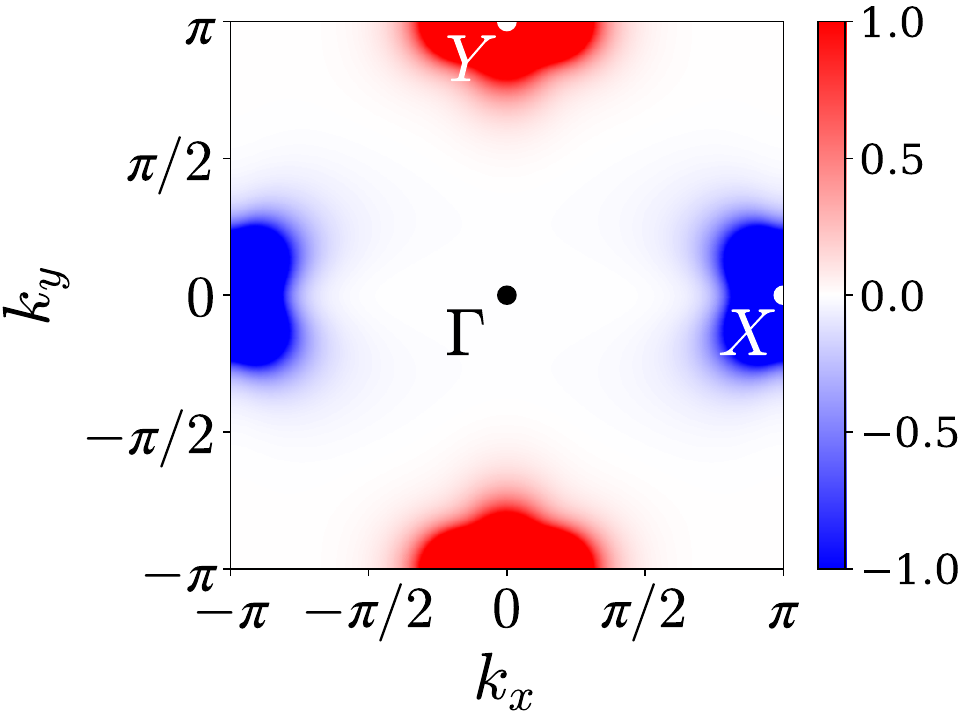}
        \label{fig:Altermagnet_color}
    } 
    \setcounter{figure}{3}
    \caption{(a) Chemical potential dependence of the SMOM at $N/N_c=0.99$ 
                 calculated by decomposing Eq.~\eqref{eq:octupole} into each term.
                 The labels ``$^{(2)}$G. P. D." and ``F. S." indicate the contributions from a two-band term proportional to the grand potential density and the Fermi sea terms, respectively.
                 The other terms are denoted by ``Others", and the total contribution is represented by ``Total"~\cite{note1}.
             (b) Momentum-resolved quantity of $\partial O_{z,xx}/\partial\mu$ at $N/N_c=0.99$.}
    \label{fig:Altermagnet_enhancement}
\end{figure}

Here, we comment on the following two points:
First, the divergent behavior of the slope at $N/N_c=1$ does not pose a problem.
This is because the slope at $N/N_c=1$ (gap-closing point) does not represent the response coefficients according to the St\v{r}eda formulas due to the absence of an insulating phase.
Second, the dominance of the two-band term, which is composed of a spin-diagonal component, is specific to this model that is block-diagonal in spin space.
More general models should involve spin-off-diagonal terms that also contribute to the SMOM.
In such models, however, the SMOM may have another interband effect and may not exhibit a clear enhancement.

\subsection{3D altermagnet model} \label{subsec:3DAltermagnet}

\begin{figure}[t]
    \setcounter{figure}{5} 
    \centering
    \subfigure[]{
        \includegraphics[height=3.5cm,width=0.46\hsize]{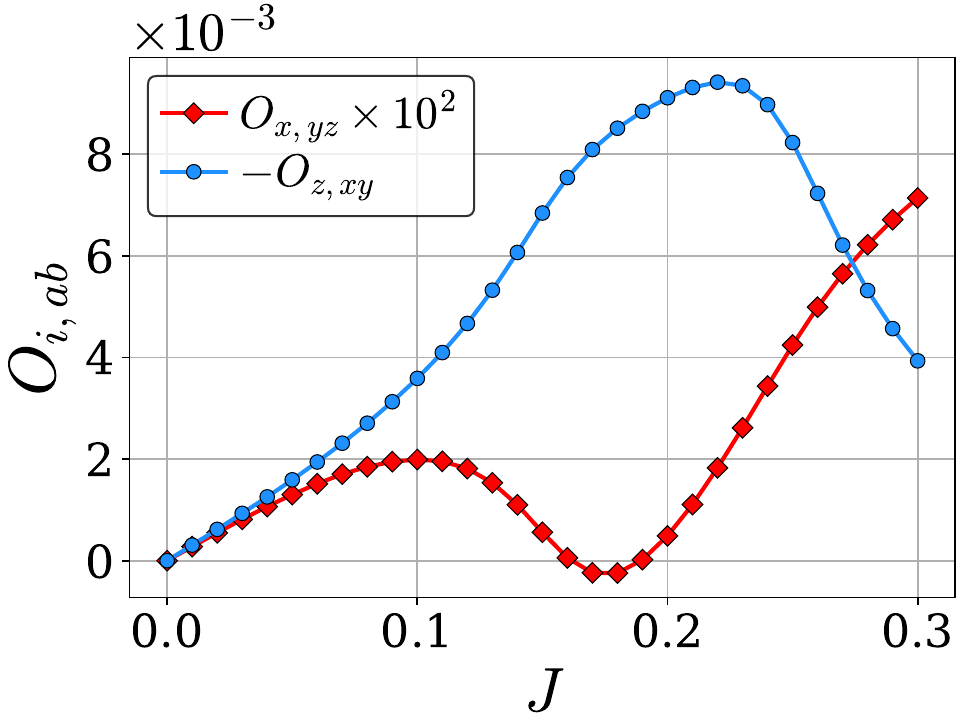}
        \label{fig:Altermagnet_Jdependence}
    }
    \hfill
    \subfigure[]{
        \includegraphics[height=3.5cm,width=0.46\hsize]{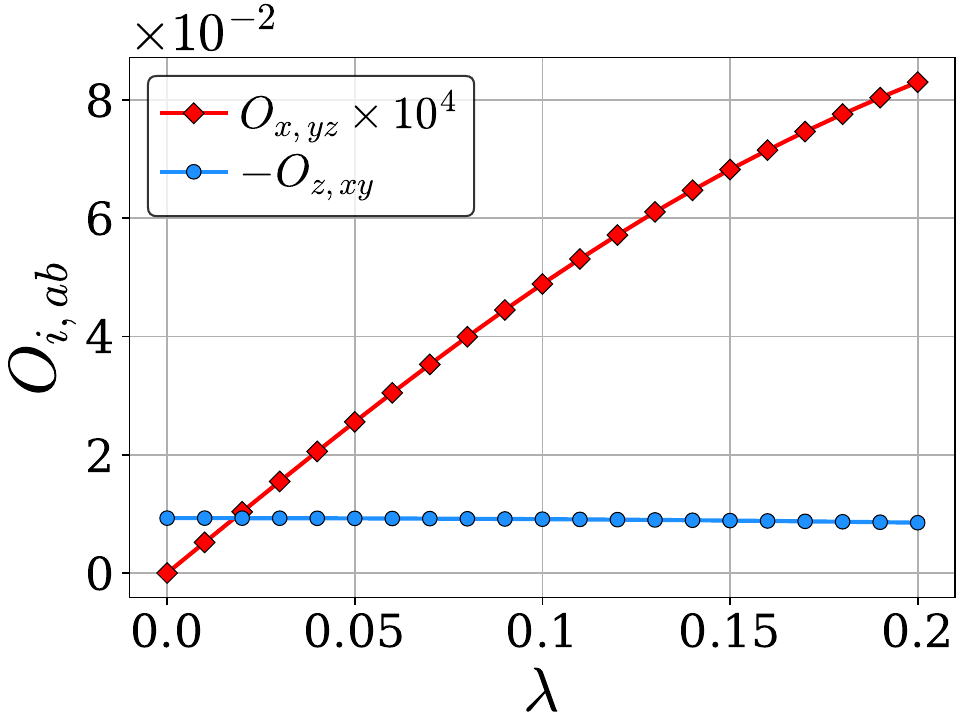}
        \label{fig:Altermagnet_SOCdependence}
    } 
    \setcounter{figure}{4}
    \caption{$J$ dependence (a) and SOC dependence (b) of the SMOM in the 3D altermagnet model.
    We set $\lambda=0.1$ in (a), $J=0.2$ in (b), and $\mu=0.25$ in both panels.}
    \label{fig:Altermagnet_3D1}
\end{figure}

\begin{figure*}
    \setcounter{figure}{6} 
    \centering
    \subfigure[$\phi=0^\circ$]{
        \includegraphics[height=4.5cm,width=0.3\hsize]{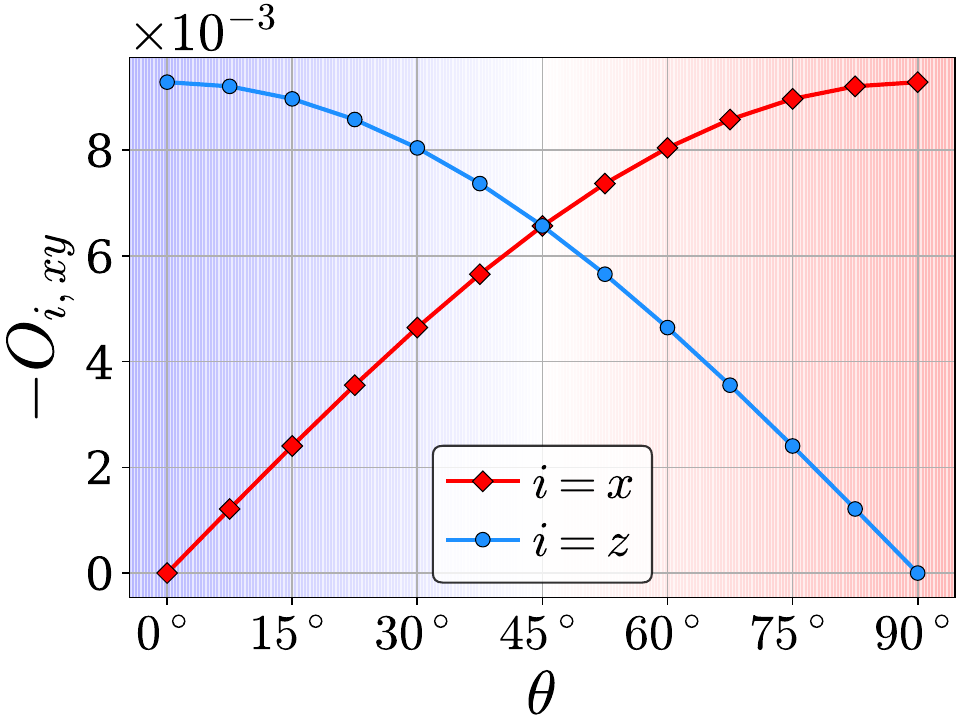}
        \label{fig:Altermagnet_thetadependence}
    }
    \hfill
    \subfigure[$\theta=90^\circ$]{
        \includegraphics[height=4.3cm,width=0.32\hsize]{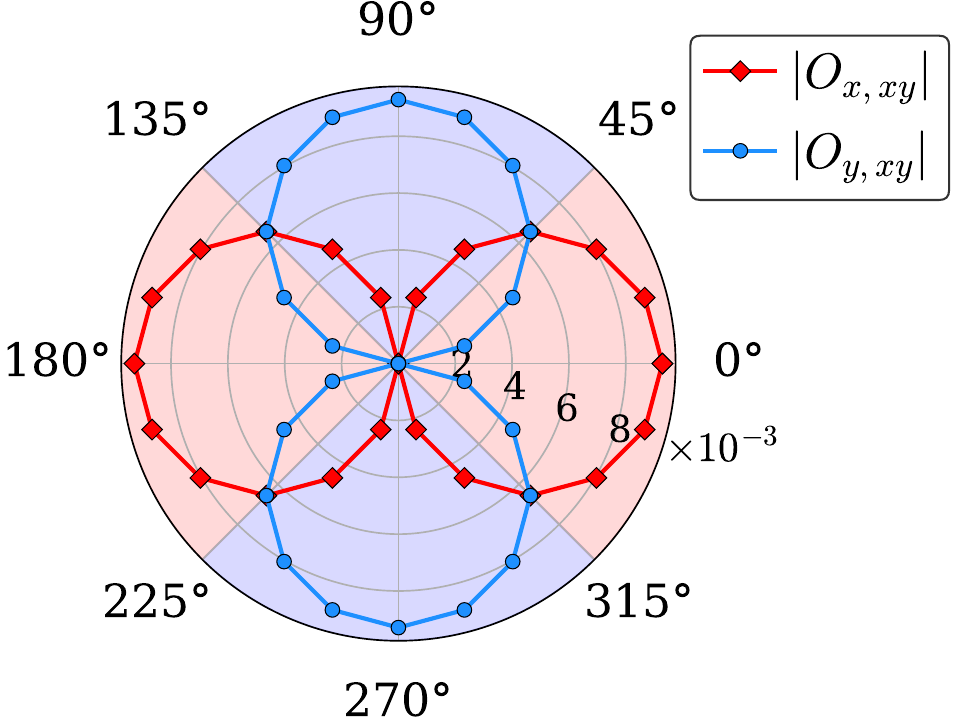}
        \label{fig:Altermagnet_phidependence}
    } 
    \hfill
    \subfigure[]{
        \includegraphics[height=4.5cm,width=0.3\hsize]{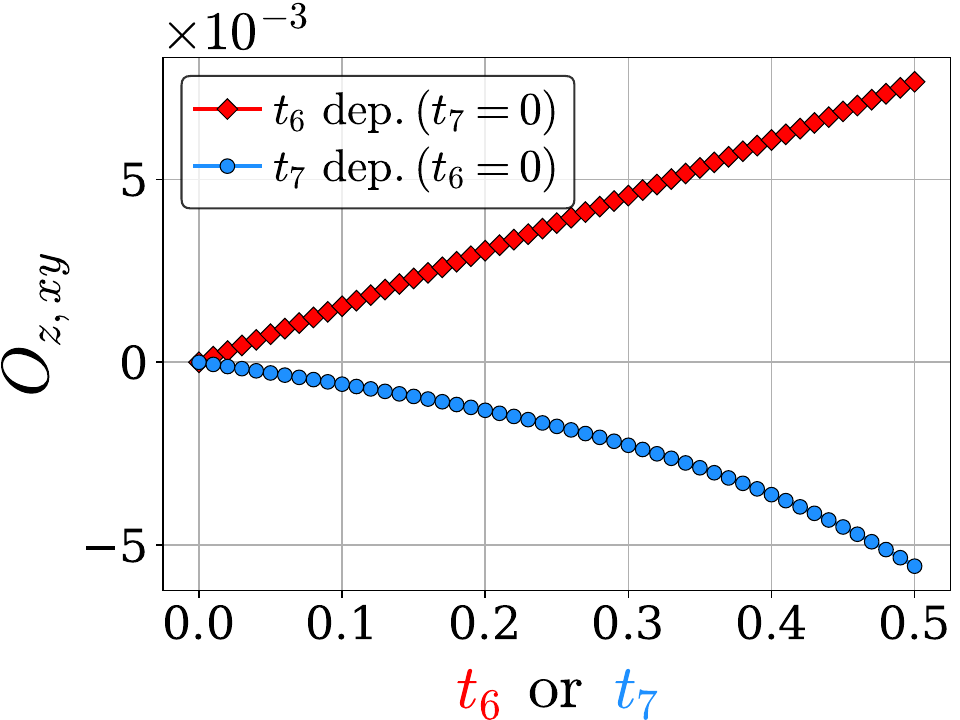}
        \label{fig:Altermagnet_t6t7dependence}
    } 
    \setcounter{figure}{5}
    \caption{(a,b) Angular dependence of the nonrelativistic SMOMs.
             We calculate the $\theta$ dependence with $\phi=0^\circ$ in (a) and $\phi$ dependence with $\theta=90^\circ$ in (b), setting $\lambda=0.0$, $J=0.2$, and $\mu=0.25$.
             In particular, (b) shows a polar plot of the absolute values.
             The red- and blue-shaded regions denote areas where the $x$ and $z\ (y)$ components are dominant, respectively.
             The dominant region changes across $45^\circ$ and its three equivalent directions.
            (c) $t_6$ and $t_7$ dependences of the nonrelativistic SMOM $O_{z,xy}$ at $\lambda=0.0$, $J=0.2$, and $\mu=0.25$.
            The red and blue lines represent the dependences on $t_6$ with $t_7=0$ and on $t_7$ with $t_6=0$, respectively. }
    \label{fig:Altermagnet_3D2}
\end{figure*}

Finally, we study the microscopic properties of the SMOM in the 3D altermagnet model.
Figure~\ref{fig:Altermagnet_Jdependence} shows how the SMOM depends on the coupling constant $J$.
The SMOM does not appear without the N\'{e}el order.
Once the N\'{e}el order develops, the SMOM initially exhibits a proportional dependence on $J$ up to a certain threshold, beyond which it begins to show a more complex dependence.
Notably, $O_{z,xy}$ is much larger than $O_{x,yz}$, which can be understood from their distinct SOC dependences, as shown in Fig.~\ref{fig:Altermagnet_SOCdependence}.
The absence of the SOC confines the spin to the N\'{e}el vector direction, resulting in a finite $O_{z,xy}$ alone.
On the other hand, the presence of the SOC induces in-plane spin components via a spin canting, leading to a finite $O_{x,yz}$.
In this context, $O_{z,xy}$ originates from nonrelativistic effects, whereas $O_{x,yz}$ stems from relativistic effects, indicating that they can be regarded as nonrelativistic and relativistic SMOMs, respectively.
Therefore, it can be rephrased that in $d$-wave altermagnets, the nonrelativistic SMOM is larger than the relativistic counterpart.
The emergence and dominance of these nonrelativistic SMOMs are absent in conventional heavy fermion systems, where magnetic multipoles are typically driven by SOC~\cite{Kuramoto2009-ln,Santini2009-er,Mydosh2020-gm}.

Next, we compare the result in Fig.~\ref{fig:Altermagnet_3D1} with the results of Landau theory and group theory for $d$-wave altermagnetism.
First, the Landau theory explains the emergence of $O_{z,xy}$ in the absence of SOC~\cite{McClarty2024-eo}.
Specifically, the free energy for the N\'{e}el vector $\bm{N}$ in the $D_{4h}$ point group has a linear coupling term $\bm{N}\cdot \bm{O}_{xy}=N_iO_{i,xy}$ in the absence of SOC, inducing $O_{z,xy}$ through the out-of-plane N\'{e}el vector $N_z$.
Then, the group theory demonstrates that the presence of SOC introduces $O_{x,yz}$ as an additional order parameter alongside $O_{z,xy}$~\cite{Fernandes2023-cw}.
Figure~\ref{fig:Altermagnet_3D1} is indeed consistent with these symmetry-based theoretical arguments, but it goes further by clarifying which component is dominant.
Specifically, it reveals that the primary contribution comes from N\'{e}el order, with SOC playing only a secondary role; consequently, $O_{z,xy}$ in Fig.~\ref{fig:Altermagnet_SOCdependence} apparently exhibits little dependence on the SOC.

To further understand the relationship between the N\'{e}el vector $\bm{N}$ and nonrelativistic SMOM $O_{i,xy}$, we focus on the linear coupling term $N_iO_{i,xy}$, which suggests that $O_{i,xy}$ strongly depends on the direction of $\bm{N}$.
Therefore, we parametrize the direction of $\bm{N}$ using angles $\theta$ and $\phi$ as $\bm{N}=(\sin\theta\cos\phi,\sin\theta\sin\phi,\cos\theta)$ and explore the angular dependence of $O_{i,xy}$.
Figures~\ref{fig:Altermagnet_thetadependence} and \ref{fig:Altermagnet_phidependence} show the $\theta$ dependence ($\phi=0^\circ$) and $\phi$ dependence ($\theta=90^\circ$) of $O_{i,xy}$, respectively.
In both cases, the component aligned with the direction of $\bm{N}$ takes the maximum value and decreases as it moves away from this direction, which embodies Landau theory for $d$-wave altermagnetism.

Furthermore, we identify the essential parameters for the nonrelativistic SMOM $O_{z,xy}$, emphasizing its relevance to the nonrelativistic spin splitting (NRSS), which is expressed as [Appendix~\ref{subsec:model_detail}]
\begin{align}
    \Delta E_s=\sqrt{t_x^2+(t_z+J)^2}-\sqrt{t_x^2+(t_z-J)^2}.
\end{align}
This expression indicates that the product $t_zJ$ induces the NRSS, with $J$ also playing a crucial role for $O_{z,xy}$, as demonstrated in Fig.~\ref{fig:Altermagnet_Jdependence}. 
We thus clarify the dependence of $O_{z,xy}$ on the intra-sublattice hopping,
\begin{align}
    t_z&=t_6 \sin k_x \sin k_y+t_7\sin k_x \sin k_y \cos k_z,
\end{align}
which arises from the distinct arrangements of nonmagnetic atoms surrounding the two sublattices~\cite{Antonenko2025-jp}.
Figure~\ref{fig:Altermagnet_t6t7dependence} shows the $t_6$ and $t_7$ dependences of $O_{z,xy}$, revealing that $t_z$ is also required for the emergence of $O_{z,xy}$.
Therefore, nonrelativistic SMOMs originate from both the distinct crystallographic environments of the sublattices ($t_z$) and N\'{e}el order ($J$), which coincides with the conditions for the emergence of NRSS.

\section{Summary} \label{sec:conclusions}

In this paper, we have derived the SMOM for bulk crystals based on a thermodynamic relation.
The obtained formula is gauge-invariant, applicable to both insulators and metals, and suitable for implementation in first-principles calculations, thereby enabling direct comparison with experimental results.
Additionally, the formula is anticipated to be useful for calculating thermal responses, as they frequently require multipole corrections to their response coefficients~\cite{Smrcka1977-ox,Cooper1997-td,Qin2011-af,Shitade2022-qy}.
For example, the magnetic quadrupole moment provides a multipole correction in the gravito-magnetoelectric effect~\cite{Shitade2019-hq,Shinada2023-pf}, where magnetization is induced by a temperature gradient; accordingly, the SMOM is expected to play a similar role.

Furthermore, we examined the relationships between the SMOM and quantum geometry in a pyrochlore model and the St\v{r}eda formulas and topology in a 2D $d$-wave altermagnet model.
Although the SMOM includes terms related to the quantum metric, we did not find an enhancement of the SMOM near Weyl points.
On the other hand, the St\v{r}eda formulas can be enhanced by interband effects in spin-conserved insulators.
Finally, studies on a 3D $d$-wave altermagnet model have confirmed the dominance of nonrelativistic SMOMs and the validity of their Landau theory~\cite{McClarty2024-eo}.

Notably, nonrelativistic SMOMs in $d$-wave altermagnets have the same microscopic origin as the NRSS that characterizes $d$-wave altermagnetism.
This common origin implies that SMOM measurements can offer an alternative approach for detecting NRSS, complementing the widely used angle-resolved photoemission spectroscopy~\cite{Fedchenko2024-lw,Lin2024-qk}.
However, detecting the SMOM remains a significant challenge due to the lack of an external field that directly couples to it.
One possible experimental approach to this problem is neutron scattering, which has been demonstrated to be effective for detecting local higher-rank magnetic multipoles~\cite{Urru2023-gy}.
Another promising method is using the response phenomena according to the St\v{r}eda formulas, which complement neutron scattering by enabling access to the bulk SMOM, i.e., magnetic octupole order.
In this context, candidate materials with magnetic octupole orders are, for example, listed in Ref.~\cite{Oike2024-bl}.

\textit{Note added}.
After completing this work and shortly before submission, we became aware of related calculations being carried out independently by another group~\cite{sato2025-rt}.
They also independently derived the SMOM consistent with our formulation, complementing our study by focusing on different irreducible components.
Specifically, our focus is on a pure SMOM devoid of any dipole components, whereas theirs is on a mixed SMOM whose dipole component is argued to characterize antiferromagnets exhibiting the anomalous Hall effect.

\section*{Acknowledgement}

We greatly thank Takumi Sato and Satoru Hayami for sharing their findings with us.
In particular, their deep insights helped us refine the description of the expansion around degenerate points, leading to an improvement in this work.
R.P. is supported by JSPS KAKENHI Grant No. JP23K03300.
K.S. acknowledges support as a JSPS research fellow and is supported by JSPS KAKENHI Grants No. 22J23393 and No. 22KJ2008. 

\section*{Data availability}
The data that support the findings of this article are available from the authors upon reasonable request.

\appendix
\renewcommand{\appendixname}{APPENDIX}

\begin{widetext}

\section{DERIVATION OF THE SMOM} \label{sec:derivation}

In this section, we derive the SMOM by expanding the density-spin correlation function,
\begin{equation}
    \Phi_i(\bm{q}) = \sum_{nm\bm{k}} \frac{f_{n\bm{k}-\bm{q}/2} - f_{m\bm{k}+\bm{q}/2} }{ \epsilon_{n\bm{k}-\bm{q}/2} - \epsilon_{m\bm{k}+\bm{q}/2} }
    \braket{u_{n\bm{k}-\bm{q}/2} | u_{m\bm{k}+\bm{q}/2} } 
    \bra{u_{m\bm{k}+\bm{q}/2}} \hat{s}^i \ket{u_{n\bm{k}-\bm{q}/2}},
\end{equation}
up to the second order in momentum $\bm{q}$ and integrating it over the chemical potential $\mu$.
Note that this correlation function satisfies $\Phi_i^*(\bm{q}) = \Phi_i(-\bm{q})$, implying that the second-order expansion coefficient is real.
In the following, we first expand this correlation function by separating it into the interband case ($n \neq m$) in Sec.~\ref{appendix:derivation:interband} and the intraband case ($n = m$) in Sec.~\ref{appendix:derivation:intraband}.
Then, we derive the final expression by integrating over $\mu$ in Sec.~\ref{appendix:derivation:final}.

\subsection{Interband case ($n\neq m$)}
\label{appendix:derivation:interband}
Expanding each component in $\bm{q}$ up to the second order, we obtain
\begin{subequations}
    \begin{gather}
        \braket{u_{n\bm{k}-\bm{q}/2} | u_{m\bm{k}+\bm{q}/2} } \simeq
        q_a \braket{n | \partial_a m} - \frac{q_a q_b}{2} \braket{\partial_a n | \partial_b m}, \\
        \bra{u_{m\bm{k}+\bm{q}/2}} \hat{s}^i \ket{u_{n\bm{k}-\bm{q}/2}}
        \simeq
        s^i_{mn} + \frac{q_a}{2} ( \bra{\partial_a m} \hat{s}^i \ket{n} - \bra{ m} \hat{s}^i \ket{\partial_a n}), \\
        \frac{f_{n\bm{k}-\bm{q}/2} - f_{m\bm{k}+\bm{q}/2} }{ \epsilon_{n\bm{k}-\bm{q}/2} - \epsilon_{m\bm{k}+\bm{q}/2} }
        \simeq
        \frac{f_{nm}}{\epsilon_{nm}} - \frac{q_a}{ 2\epsilon_{nm}}
        \Bigl(
        \partial_a \tilde{f}_{nm} - \frac{\partial_a \tilde{\epsilon}_{nm} f_{nm}}{\epsilon_{nm}}
        \Bigr).
    \end{gather}
\end{subequations}
Here, we apply the Einstein summation convention for repeated indices and use the following abbreviations:
\begin{gather*}
    \ket{n} = \ket{u_{n\bm{k}}},~~ \partial_a = \partial/\partial k_a,~~M_{nm} = \bra{n} \hat{M} \ket{m},~~f_{nm} = f_n - f_m,~~\epsilon_{nm} = \epsilon_{n\bm{k}} - \epsilon_{m\bm{k}}, \nonumber \\
    \tilde{f}_{nm} = f_n + f_m,~~
    \tilde{\epsilon}_{nm} = \epsilon_{n\bm{k}} + \epsilon_{m\bm{k}}.
\end{gather*}
Using these equations, the second-order term of the correlation function
becomes
\begin{align}
    \Phi^{(\mathrm{A})}_i =& \sum_{n \neq m \bm{k}}q_a q_b \mathrm{Re}
    \Bigl[
    -\frac{1}{2} \braket{\partial_a n | \partial_b m} s^i_{mn} \frac{f_{nm}}{\epsilon_{nm}}
    +\braket{n | \partial_a m} \frac{1}{2}(\bra{\partial_b m} \hat{s}^i \ket{n} - \bra{ m} \hat{s}^i \ket{\partial_b n}) ) \frac{f_{nm}}{\epsilon_{nm}} \nonumber \\
    & + \braket{n | \partial_a m} s^i_{mn} \left(\frac{-1}{2\epsilon_{nm}}\right) \Bigl(
        \partial_b \tilde{f}_{nm} - \frac{\partial_b \tilde{\epsilon}_{nm} f_{nm}}{\epsilon_{nm}}
        \Bigr)
    \Bigr].
\end{align}
This $q$-expansion includes derivatives of the wave function, seemingly breaking gauge invariance.
However, this gauge-invariance breaking is artificial, and the correlation function remains gauge-invariant.
To see this fact, we replace the derivative $\partial_a$ with the covariant derivative $D_a$ and diagonal Berry connection $A^a_n$ based on Eq.~(\ref{eq:covariant_derivative}).
The terms including the diagonal Berry connection are 
\begin{equation}
    \sum_{n \neq m \bm{k}} q_a q_b \mathrm{Re} \Bigl[
        -\frac{1}{2}( \braket{D_a n | m} (- iA_m^b) + \braket{ n |D_b m} ( iA_n^a) )
         s^i_{mn} \frac{f_{nm}}{\epsilon_{nm}} + \frac{1}{2} \braket{ n |D_a m} ((i A_m^b + iA_n^b) s^i_{mn} ) \frac{f_{nm}}{\epsilon_{nm}} 
    \Bigr] = 0.
\end{equation}
Thus, the correlation function $\Phi^{(\mathrm{A})}_i$ is rewritten by using only the covariant derivative as
\begin{align}
    \Phi^{(\mathrm{A})}_i =& \sum_{n \neq m \bm{k}}q_a q_b \mathrm{Re}
    \Bigl[
    -\frac{1}{2} \braket{D_a n | D_b m} s^i_{mn} \frac{f_{nm}}{\epsilon_{nm}}
    +\braket{n | D_a m} \frac{1}{2}(\bra{D_b m} \hat{s}^i \ket{n} - \bra{ m} \hat{s}^i \ket{D_b n}) ) \frac{f_{nm}}{\epsilon_{nm}} \nonumber \\
    & + \braket{n | D_a m} s^i_{mn} \frac{-1}{2\epsilon_{nm}} \Bigl(
        \partial_b \tilde{f}_{nm} - \frac{\partial_b \tilde{\epsilon}_{nm} f_{nm}}{\epsilon_{nm}}
        \Bigr)
    \Bigr], \label{eq:phiA}
\end{align}
which confirms gauge invariance.

\subsection{Intraband case ($n=m$)}
\label{appendix:derivation:intraband}
Similarly, expanding each component in $\bm{q}$ up to the second order, we obtain
\begin{subequations}
\begin{gather}
    \braket{u_{n\bm{k}-\bm{q}/2} | u_{n\bm{k}+\bm{q}/2}}
    \simeq
    1 + q_a \braket{ n | \partial_a n } - \frac{q_a q_b}{2} \braket{ \partial_a n | \partial_b n }, \\
    \bra{u_{n\bm{k}+\bm{q}/2}} \hat{s}^i \ket{u_{n\bm{k}-\bm{q}/2}}
    \simeq
    s^i_{n} + \frac{q_a}{2} ( \bra{\partial_a n} \hat{s}^i \ket{n} - \bra{ n} \hat{s}^i \ket{\partial_a n}) \nonumber  \\
    \hspace{120pt}+\frac{q_aq_b}{4} ( -2 \bra{\partial_a n} \hat{s}^i \ket{\partial_b n}  + \frac{1}{2} \partial_{ab} s^i_{n}  ), \\
    \frac{f_{n\bm{k}-\bm{q}/2} - f_{n\bm{k}+\bm{q}/2} }{ \epsilon_{n\bm{k}-\bm{q}/2} - \epsilon_{n\bm{k}+\bm{q}/2} }
    \simeq
    f'_n + q_aq_b \Bigl( \frac{1}{24} f'''_n v^a_{n} v^b_{n} + \frac{1}{8} \partial_{ab} \epsilon_n f''_n \Bigr),
\end{gather}
\end{subequations}
where $\partial_{ab}=\partial_a\partial_b$ and $v^a_n=\partial_a \epsilon_{n\bm{k}}$.
Thus, the second-order term of the correlation function becomes
\begin{align}
    \Phi^{(\mathrm{B})}_i =&
    \sum_{n \bm{k}} q_a q_b
    \mathrm{Re} \Bigl[
        \frac{f'_n}{4} ( -2 \bra{\partial_a n} \hat{s}^i \ket{\partial_b n}  + \frac{1}{2} \partial_{ab} s^i_{n}  )
        + \frac{f'_n}{2} \braket{n | \partial_a n} ( \bra{\partial_b n} \hat{s}^i \ket{n} - \bra{ n} \hat{s}^i \ket{\partial_b n}) \nonumber \\
    & -\frac{f'_n}{2} \braket{\partial_a n | \partial_b n} s^i_{n}    
    +
    \Bigl( \frac{1}{24} f'''_n v^a_{n} v^b_{n} + \frac{1}{8} \partial_{ab} \epsilon_n f''_n \Bigr) s^i_{n}
    \Bigr].
\end{align}
Similar to the above discussion, we can replace the derivative $\partial_a$ with the covariant derivative $D_a$ and diagonal Berry connection $A^a_n$, resulting in the cancellation of the terms that depend on the diagonal Berry connection: 
\begin{align}
    &\sum_{n\bm{k}} q_a q_b \mathrm{Re} \Bigl[
        -\frac{f'_n}{2} (\bra{D_a n } \hat{s}^i \ket{n} (-i A_n^b) + \bra{n } \hat{s}^i \ket{ D_a n} (i A_n^b)
        -s^i_{n} (iA_n^a) (- iA_n^b)
        )
        \nonumber \\
    &+\frac{f'_n}{2} (-i A_n^a) ( ( i A^b_n) (s^i_{n} - s^i_{n}) +   \bra{D_b n} \hat{s}^i \ket{n} - \bra{ n} \hat{s}^i \ket{D_b n}) 
    - \frac{f'_n}{2} (i A^a_n ) (- iA^b_n) s^i_{n}
    \Bigr] = 0.
\end{align}
Thus, the correlation function $\Phi^{(\mathrm{B})}_i$ is also rewritten by using only the covariant derivative as
\begin{align}
    \Phi^{(\mathrm{B})}_i =&
    \sum_{n \bm{k}} q_a q_b
    \mathrm{Re} \Bigl[
        \frac{f'_n}{4} ( -2 \bra{D_a n} \hat{s}^i \ket{D_b n}  + \frac{1}{2} \partial_{ab} s^i_{n}  )   
    -\frac{f'_n}{2} \braket{D_a n | D_b n} s^i_{n} 
    +
    \Bigl( \frac{1}{24} f'''_n v^a_{n} v^b_{n} + \frac{1}{8} \partial_{ab} \epsilon_n f''_n \Bigr) s^i_{n}
    \Bigr] \nonumber \\
    =&
    \sum_{n \bm{k}} q_a q_b
    \mathrm{Re} \Bigl[ - \frac{f'_n}{2} ( \bra{D_a n} \hat{s}^i \ket{D_b n} +  \braket{D_a n | D_b n} s^i_{n})
    - \frac{f''_n}{4} v^b_n \partial_a s^i_n
    - \frac{f'''_n}{12} v^a_n v^b_n s^i_n
    \Bigr]. \label{eq:phiB}
\end{align}
Here, $\partial_as^i_n$ can be transformed as
\begin{equation}
    \partial_a s^i_n = \bra{\partial_a n} \hat{s}^i \ket{n} + \bra{n} \hat{s}^i \ket{\partial_a n} = \bra{D_a n} \hat{s}^i \ket{n} + \bra{n} \hat{s}^i \ket{D_a n}.
\end{equation}

\subsection{Final expression} \label{appendix:derivation:final}

Finally, the SMOM is obtained by integrating over the chemical potential $\mu$ as 
\begin{align}
    O_{i,ab}
    =&
    \frac{1}{2}\sum_{n \bm{k}}  
    \mathrm{Re} \biggl[
    \sum_{m(\neq n)}
    \frac{\mathcal{G}_{nm}}{2} \biggl(
    \frac{ \braket{ D_a n  | D_b m} s^i_{mn} }{ \epsilon_{nm}}
    -
    \frac{ \braket{ n  | D_a m} (\braket{D_b m | \hat{s}^i | n} - \braket{ m | \hat{s}^i | D_b n}  ) }{ \epsilon_{nm}}
    \biggr) \notag \\
    &+\sum_{m(\neq n)}
    \frac{1}{2}\biggl(
    \partial_b \tilde{\mathcal{G}}_{nm} - \frac{\partial_b \tilde{\epsilon}_{nm} \mathcal{G}_{nm} }{ \epsilon_{nm}}
    \biggr)\frac{ \braket{n | D_a m} s^i_{mn}}{\epsilon_{nm}} + \frac{f_n}{2} ( \bra{D_a n} \hat{s}^i \ket{D_b n} +  \braket{D_a n | D_b n} s^i_{n}) \notag \\
    & +\frac{f'_n}{4} v^b_n (\bra{D_a n} \hat{s}^i \ket{n} + \bra{n} \hat{s}^i \ket{D_a n})
    +
    \frac{f''_n}{12} v^a_n v^b_n s^i_n
    \biggr] + (a \leftrightarrow b), \label{eq:octupole_finalexpression}
\end{align}
where $\mathcal{G}_{nm}=\mathcal{G}_n-\mathcal{G}_m$ and $\tilde{\mathcal{G}}_{nm}=\mathcal{G}_n+\mathcal{G}_m$.
In this process, we use the following formulas:
\begin{equation}
    \frac{\partial{\mathcal{G}_n}}{\partial \mu } = -f_n,~~
    \frac{\partial f^{(i)}_n}{\partial \mu} = - f^{(i+1)}_n.
\end{equation}

\section{DETAILS OF THE MODELS AND THE SYMMETRY CONSTRAINTS OF THE SMOM} \label{sec:model_detail}

\subsection{Details of the altermagnet models} \label{subsec:model_detail}

The altermagnet models for the numerical calculations read
\begin{align}
    H(\bm{k})=\epsilon_0+t_x\tau_x+t_z\tau_z+\tau_y \bm{\lambda}\cdot\bm{\sigma}+\tau_z\bm{J}\cdot\bm{\sigma}, \label{eq:altermagnet_models}
\end{align}
where $\bm{\lambda}=\lambda\bm{g}$ and $\bm{J}=J\bm{N}$.
For the 2D altermagnet model with $\bm{g}=(0,0,g_z)$ and $\bm{N}=(0,0,N)$, each coefficient of Eq.~\eqref{eq:altermagnet_models} is described as~\cite{Antonenko2025-jp}
\begin{align}
    \epsilon_0&=-2t_1(\cos k_x+\cos k_y), \quad t_x=-4t_2\cos \frac{k_x}{2}\cos \frac{k_y}{2}, \quad t_z=-2t_d(\cos k_x-\cos k_y), \quad g_z = \sin \frac{k_x}{2}\sin\frac{k_y}{2}.
\end{align}
Similarly, each coefficient of Eq.~\eqref{eq:altermagnet_models} for the 3D altermagnet model with $\bm{g}=(g_x,g_y,0)$ and $\bm{N}=(0,0,1)$ is expressed as~\cite{Roig2024-zg}
\begin{align}
    \begin{aligned}
        \epsilon_0&=t_1(\cos k_x+\cos k_y)+t_2 \cos k_z +t_3 \cos k_x \cos k_y +t_4(\cos k_x+\cos k_y)\cos k_z +t_5\cos k_x \cos k_y \cos k_z, \\
        t_x&=t_8\cos \frac{k_x}{2}\cos \frac{k_y}{2} \cos \frac{k_z}{2}, \quad t_z=t_6 \sin k_x \sin k_y + t_7\sin k_x \sin k_y \cos k_z, \quad g_{x/y} = \pm \sin \frac{k_z}{2}\sin\frac{k_{x/y}}{2} \cos \frac{k_{y/x}}{2}, 
        \end{aligned}
\end{align}
where $g_x$ and $g_y$ take the upper ($+$) and lower ($-$) signs of $\pm$, respectively.
In the numerical calculations, we use the following parameters unless otherwise noted: $t_1=0.5t_2$, $t_2=0.1$, $t_d=2t_2$, $\lambda=2t_2$, and $J=0.1$ for the 2D model and $t_1=-0.05$, $t_2=0.7$, $t_3=0.5$, $t_4=-0.15$, $t_5=-0.4$, $t_6=-0.6$, $t_7=0.3$, and $t_8=1.7$ for the 3D model.

Here, we analyze the energy dispersion of a block-diagonal Hamiltonian in spin space, which is trivial in the 2D model and corresponds to the absence of SOC in the 3D model.
In this case, the energy eigenvalues of the up-spin ($\uparrow$) and down-spin ($\downarrow$) bands are given by
\begin{align}
    E^{\uparrow}_{\pm}=\epsilon_0 \pm \sqrt{ t_x^2+(t_z+J_z)^2+\lambda_z^2}, \quad  E^{\downarrow}_{\pm}=\epsilon_0 \pm \sqrt{ t_x^2+(t_z-J_z)^2+\lambda_z^2}, \label{eq:altermagnet_models_energy}
\end{align}
where $\lambda_z=0$ in the 3D model.
In the 2D model, the coefficients become $\epsilon_0=t_x=\lambda_z=0$ and $t_z=4t_d$ at the $X(\pi,0)$ point and $\epsilon_0=t_x=\lambda_z=0$ and $t_z=-4t_d$ at the $Y(0,\pi)$ point, reducing Eq.~\eqref{eq:altermagnet_models_energy} to Eq.~\eqref{eq:Altermagnet_models_energy2D_XY} in the main text.
Furthermore, Eq.~\eqref{eq:altermagnet_models_energy} describes spin splitting as $\Delta E_s=E^{\uparrow}_{\pm}-E^{\downarrow}_{\pm}$.

\begin{table}[t]
    \caption{Symmetry constraints of the SMOM in the pyrochlore and 2D/3D altermagnet models. 
    The symbol ``$\checkmark/\times$" denotes whether each component is allowed ($\checkmark$) or forbidden ($\times$) under the symmetry operation.
    Note that we use the relation $O_{i,ab}=O_{i,ba}$.}
    \label{tab:Symmetry_Constraints}
    \centering
    \tabcolsep = 2.5mm
    \renewcommand\arraystretch{1.4}
    \begin{tabular}{cccccccccc} \hline \hline
    & $\mathcal{P}$ & $C^{[111]}_{3}$ & $C^{[110]}_{2}\mathcal{T}$ & $C^{[001]}_{4}\mathcal{T}$ & $\mathcal{M}_{(100)}$ & $\mathcal{M}_{(010)}$  & $\mathcal{M}_{(001)}$ & $\mathcal{M}_{(110)} $ & $\mathcal{M}_{(100)} \mathcal{T}$  \\ \hline 
    $O_{x,xx}$ & $\checkmark$ & $\checkmark$ & $\checkmark$ & $\times$ & $\checkmark$ & $\times$ & $\times$ & $\checkmark$ & $\times$  \\ 
    $O_{x,yy}$ & $\checkmark$ & $\checkmark$ & $\checkmark$ & $\times$ & $\checkmark$ & $\times$ &  $\times$ &  $\checkmark$ & $\times$  \\
    $O_{x,zz}$ & $\checkmark$ & $\checkmark$ & $\checkmark$ & $\times$ & $\checkmark$ & $\times$ &  $\times$ &  $\checkmark$ & $\times$  \\
    $O_{x,yz}$ & $\checkmark$ & $\checkmark$ &  $\checkmark$ & $\checkmark$ & $\checkmark$ & $\checkmark$ & $\checkmark$ & $\checkmark$ & $\times$  \\
    $O_{x,zx}$ & $\checkmark$ & $\checkmark$ & $\checkmark$ & $\checkmark$ & $\times$ & $\times$ & $\checkmark$ & $\checkmark$ & $\checkmark$  \\
    $O_{x,xy}$ & $\checkmark$ & $\checkmark$ & $\checkmark$ & $\times$ & $\times$ &  $\checkmark$ & $\times$ &  $\checkmark$ & $\checkmark$  \\ \hline
    $O_{y,xx}$ & $\checkmark$ & $O_{x,zz}$ & $-O_{x,yy}$ & $\times$ & $\times$ & $\checkmark$ & $\times$ & $O_{x,yy}$ & $\checkmark$  \\ 
    $O_{y,yy}$ & $\checkmark$ & $O_{x,xx}$ & $-O_{x,xx}$ & $\times$ & $\times$ &  $\checkmark$ &  $\times$ &  $O_{x,xx}$ & $\checkmark$  \\
    $O_{y,zz}$ & $\checkmark$ & $O_{x,yy}$ & $-O_{x,zz}$ & $\times$ &  $\times$ &  $\checkmark$ &  $\times$ &  $O_{x,zz}$ & $\checkmark$  \\
    $O_{y,yz}$ & $\checkmark$ & $O_{x,xy}$ & $O_{x,zx}$ & $-O_{x,zx}$ & $\times$ & $\times$ & $\checkmark$ & $-O_{x,zx}$ & $\checkmark$  \\
    $O_{y,zx}$ & $\checkmark$ &  $O_{x,yz}$ & $O_{x,yz}$ & $O_{x,yz}$ & $\checkmark$ & $\checkmark$ & $\checkmark$ & $-O_{x,yz}$ &  $\times$ \\
    $O_{y,xy}$ & $\checkmark$ & $O_{x,zx}$ & $-O_{x,xy}$ & $\times$ & $\checkmark$ & $\times$ &  $\times$ &  $O_{x,xy}$ & $\times$  \\ \hline
    $O_{z,xx}$ & $\checkmark$ & $O_{y,zz}$ & $\checkmark$ & $\checkmark$ & $\times$ & $\times$ & $\checkmark$ & $\checkmark$ & $\checkmark$  \\ 
    $O_{z,yy}$ & $\checkmark$ & $O_{y,xx}$ & $O_{z,xx}$ & $-O_{z,xx}$ & $\times$ & $\times$ & $\checkmark$ & $-O_{z,xx}$ & $\checkmark$  \\
    $O_{z,zz}$ & $\checkmark$ & $O_{y,yy}$ & $\checkmark$ & $\times$ & $\times$ & $\times$ & $\checkmark$ & $\times$ & $\checkmark$  \\
    $O_{z,yz}$ & $\checkmark$ & $O_{y,xy}$ & $\checkmark$ & $\times$ & $\times$ & $\checkmark$ &  $\times$ &  $\checkmark$ & $\checkmark$  \\
    $O_{z,zx}$ & $\checkmark$ & $O_{y,yz}$ & $-O_{z,yz}$ & $\times$ & $\checkmark$ & $\times$ &  $\times$ &  $O_{z,yz}$ & $\times$ \\
    $O_{z,xy}$ & $\checkmark$ &  $O_{y,zx}$ &  $\checkmark$ &  $\checkmark$ & $\checkmark$ & $\checkmark$ & $\checkmark$ & $\times$ & $\times$  \\ \hline \hline
    \end{tabular}
\end{table}

\subsection{Symmetry constraints of the SMOM} \label{subsec:symmetry_detail}

In this subsection, we derive the symmetry constraints of the SMOM in the pyrochlore and 2D/3D altermagnet models.
These models belong to $m\bar{3}m'$ and $4'/mm'm$, respectively, with the following symmetry operations:
\begin{align}
    m\bar{3}m'&=\{ \mathcal{P}, C^{[111]}_{3}, \mathcal{M}_{(100)}, \mathcal{M}_{(001)}, C^{[110]}_2 \mathcal{T} \}, \\
    4'/mm'm&=\left\{
    \begin{aligned}
        &\{ \mathcal{P}, C^{[001]}_{4}\mathcal{T}, \mathcal{M}_{(110)},\mathcal{M}_{(100)}\mathcal{T}\} \quad (\mathrm{for}\ d_{x^2-y^2}\ \mathrm{symmetry})~\cite{Fang2023-xl} \\
        &\{ \mathcal{P}, C^{[001]}_{4}\mathcal{T}, \mathcal{M}_{(100)}, \mathcal{M}_{(010)}\} \quad (\mathrm{for}\ d_{xy}\ \mathrm{symmetry})~\cite{Xiao2022-xr}
    \end{aligned},
    \right.
\end{align}
where $\mathcal{P}$ is spatial inversion, $\mathcal{T}$ is time reversal, $C^{[abc]}_n$ is a $2\pi/n$ rotation around the $[a,b,c]$ axis, and $\mathcal{M}_{(abc)}$ is a mirror in the $(a,b,c)$ plane.
The matrix representations of these symmetry operations in 3D space are given by 
\begin{align}
    \begin{aligned}
        &\mathcal{P}=
        \begin{pmatrix}
            1 & 0 & 0 \\
            0 & 1 & 0 \\
            0 & 0 & 1
        \end{pmatrix}, \quad 
        C^{[110]}_2=
        \begin{pmatrix}
            0 & 1 & 0 \\
            1 & 0 & 0 \\
            0 & 0 & -1
        \end{pmatrix}, \quad 
        C^{[111]}_3=
        \begin{pmatrix}
            0 & 0 & 1 \\
            1 & 0 & 0 \\
            0 & 1 & 0 
        \end{pmatrix}, \quad 
        C^{[001]}_4=
        \begin{pmatrix}
            0 & -1 & 0 \\
            1 & 0 & 0 \\
            0 & 0 & 1
        \end{pmatrix}, \\
        &\mathcal{M}_{(100)}=
        \begin{pmatrix}
            -1 & 0 & 0 \\
            0 & 1 & 0 \\
            0 & 0 & 1
        \end{pmatrix}, \quad
        \mathcal{M}_{(010)}=
        \begin{pmatrix}
            1 & 0 & 0 \\
            0 & -1 & 0 \\
            0 & 0 & 1
        \end{pmatrix}, \quad 
        \mathcal{M}_{(001)}=
        \begin{pmatrix}
            1 & 0 & 0 \\
            0 & 1 & 0 \\
            0 & 0 & -1
        \end{pmatrix}, \quad 
        \mathcal{M}_{(110)}=
        \begin{pmatrix}
            0 & -1 & 0 \\
            -1 & 0 & 0 \\
            0 & 0 & 1
        \end{pmatrix}.
    \end{aligned} \label{eq:symmetry_operations}
\end{align}
Here, we use the fact that the rotation matrix $C^{[n_1 n_2 n_3]}_{2\pi/\theta}$, which is a $\theta$ rotation around an arbitrary axis $[n_1,n_2,n_3]$, is expressed as
\begin{align}
    C^{[n_1 n_2 n_3]}_{2\pi/\theta}=
    \begin{pmatrix}
        \tilde{n}_1^2(1-\cos\theta)+\cos\theta & \tilde{n}_1 \tilde{n}_2(1-\cos\theta)-\tilde{n}_3\sin\theta & \tilde{n}_1 \tilde{n}_3(1-\cos\theta)+\tilde{n}_2\sin\theta \\
        \tilde{n}_1 \tilde{n}_2(1-\cos\theta)+\tilde{n}_3\sin\theta & \tilde{n}_2^2(1-\cos\theta)+\cos\theta & \tilde{n}_2 \tilde{n}_3(1-\cos\theta)-\tilde{n}_1\sin\theta \\
        \tilde{n}_1 \tilde{n}_3(1-\cos\theta)-\tilde{n}_2\sin\theta & \tilde{n}_2 \tilde{n}_3(1-\cos\theta)+\tilde{n}_1\sin\theta & \tilde{n}_3^2(1-\cos\theta)+\cos\theta 
    \end{pmatrix},
\end{align}
where $\tilde{n}_i=n_i/\sqrt{n_1^2+n_2^2+n_3^2}$.
Applying Eq.~\eqref{eq:symmetry_operations} to Eq.~\eqref{eq:classification}, we obtain Table~\ref{tab:Symmetry_Constraints}, which summarizes the symmetry constraints of the SMOM, leading to the descriptions in the main text.

\section{EXPRESSION OF THE SMOM FOR NUMERICAL CALCULATIONS} \label{sec:SMOM_numerical}

Applying the Hellmann-Feynman theorem $\braket{n|D_a m}=-v^a_{nm}/\epsilon_{nm}$ to Eq.~\eqref{eq:octupole_finalexpression}, we obtain the expression of the SMOM for numerical calculations as 
\begin{align}
    O_{i,ab}
    =&
    \frac{1}{2}\sum_{n\bm{k}} 
    \mathrm{Re} \biggl[
    \sum_{m(\neq n)} \mathcal{G}_{nm} \biggl(
    \underbrace{-\frac{1}{2}\sum_{l(\neq n,m)} \frac{v^a_{nl}v^b_{lm}s^i_{mn}}{\epsilon_{nl}\epsilon_{lm}\epsilon_{nm}}}_{(\mathrm{I})}+\underbrace{\sum_{l(\neq n)} \frac{v^a_{nm}s^i_{ml}v^b_{ln}}{\epsilon^2_{nm}\epsilon_{ln}}}_{(\mathrm{II})}
    \biggr)
     + \underbrace{\frac{1}{2}\sum_{m(\neq n)} \biggl(
    -\partial_b \tilde{\mathcal{G}}_{nm} + \frac{\partial_b \tilde{\epsilon}_{nm} \mathcal{G}_{nm} }{ \epsilon_{nm}}
    \biggr)\frac{v^a_{nm}s^i_{mn}}{\epsilon^2_{nm}}}_{(\mathrm{III})} \nonumber \\
    &
    \quad + \frac{f_n}{2} \biggl( \underbrace{- \sum_{m (\neq n)}\sum_{l (\neq n)}\frac{v^a_{nm}s^i_{ml}v^b_{ln}}{\epsilon_{nm}\epsilon_{ln}}}_{(\mathrm{IV})}+\underbrace{\sum_{m(\neq n)}\frac{v^a_{nm}v^b_{mn}s^i_n}{\epsilon^2_{nm}}}_{(\mathrm{V})}
     \biggr) + \underbrace{ \frac{f'_n}{2} v^b_n \sum_{m (\neq n)} \frac{v^a_{nm}s^i_{mn}}{\epsilon_{nm}}}_{(\mathrm{VI})}+
    \frac{f''_n}{12} v^a_n v^b_n s^i_n
    \biggr] + (a \leftrightarrow b). \label{eq:octupole_calculation}
\end{align}

\section{EXPRESSION OF THE SMOM NEAR DEGENERATE POINTS} \label{sec:SMOM_degeneracy}

In this section, we confirm the convergence of Eq.~\eqref{eq:octupole_calculation} at degenerate points and derive a formula applicable near them.
The detailed procedure is performed by decomposing Eq.~\eqref{eq:octupole_calculation} into two-band terms in Sec.~\ref{subsec:two-band} 
and multi-band terms in Sec.~\ref{subsec:multi-band} and then expanding them in terms of an infinitesimal quantity $\Delta \epsilon\ (\ll1)$.
Note that we use the property $(a \leftrightarrow b)$ in the following, but omit the description for simplicity.

\subsection{Two-band terms} \label{subsec:two-band}

Here, we assume $\epsilon_n=\epsilon_m$ at a degenerate point and repeatedly apply the following expansion formula:
\begin{align}
    \mathcal{G}_{n}=\mathcal{G}(\epsilon_m+\epsilon_{nm})=\mathcal{G}_m+f_m\epsilon_{nm}+\frac{f'_m}{2}\epsilon_{nm}^2+\frac{f''_m}{6}\epsilon_{nm}^3+\frac{f'''_m}{24}\epsilon_{nm}^4+\cdots. 
\end{align}
First, the two-band components in terms (II), (IV), and (V) are expanded as
\begin{align}
    \bigl\{(\mathrm{II})+(\mathrm{IV})+(\mathrm{V})\bigr\}_{\mathrm{two \textendash band}}&=\biggl( -\mathcal{G}_{nm}\frac{1}{\epsilon^3_{nm}}+\frac{f_n}{2}\frac{1}{\epsilon^2_{nm}}+\frac{f_m}{2}\frac{1}{\epsilon^2_{nm}} \biggr) \mathrm{Re}\bigl[  v^a_{nm}s^i_{m}v^b_{mn} \bigr] \notag \\
    &=\biggl( \frac{f_{nm}}{2}\frac{1}{\epsilon^2_{nm}}-\frac{f'_m}{2}\frac{1}{\epsilon_{nm}}-\frac{f''_m}{6}-\frac{f'''_m}{24}\epsilon_{nm} \biggr) \mathrm{Re}\bigl[  v^a_{nm}s^i_{m}v^b_{mn} \bigr] +\mathcal{O}(\epsilon^2_{nm})\notag \\
    &= \frac{1}{12}\biggl( f''_m+\frac{f'''_m}{2}\epsilon_{nm} \biggr) \mathrm{Re}\bigl[  v^a_{nm}s^i_{m}v^b_{mn} \bigr] +\mathcal{O}(\epsilon^2_{nm}), \label{eq:II-IV-V}
\end{align}
where we perform $n \leftrightarrow m$ in term (V).
Similarly, term (III) is transformed as
\begin{align}
    (\mathrm{III})&=\frac{1}{2}\mathrm{Re} \biggl[ \biggl\{-\partial_b\epsilon_nf_n-\partial_b\epsilon_mf_m+\partial_b\tilde{\epsilon}_{nm} \Bigl( f_m+\frac{f'_m}{2}\epsilon_{nm}+\frac{f''_m}{6}\epsilon^2_{nm}+\frac{f'''_m}{24}\epsilon^3_{nm} \Bigr)
    \biggr\} \frac{v^a_{nm}s^i_{mn}}{\epsilon^2_{nm}} \biggr]+\mathcal{O}(\epsilon^2_{nm}) \notag \\
    &=\frac{1}{2}\biggl( -f_{nm} \frac{1}{\epsilon^2_{nm}}v^b_n-\frac{f'_{nm}}{4}\frac{1}{\epsilon_{nm}}\partial_b\tilde{\epsilon}_{nm}+\frac{f''_m}{6}\partial_b\tilde{\epsilon}_{nm}+\frac{f'''_m}{24}\partial_b\tilde{\epsilon}_{nm}\epsilon_{nm}\biggr) \mathrm{Re} \bigl[ v^a_{nm}s^i_{mn} \bigr]+\mathcal{O}(\epsilon^2_{nm}) \notag \\
    &=\frac{1}{2}\biggl( -f'_m \frac{1}{\epsilon_{nm}} -\frac{f''_m}{2} -\frac{f'''_m}{6} \epsilon_{nm} \biggr)\mathrm{Re} \bigl[ v^b_n v^a_{nm}s^i_{mn} \bigr] -\frac{1}{24}(f''_m+f'''_m\epsilon_{nm} )\mathrm{Re}\bigl[ \partial_b\tilde{\epsilon}_{nm}v^a_{nm}s^i_{mn} \bigr]+\mathcal{O}(\epsilon^2_{nm}) \label{eq:III},
\end{align}
where to obtain the second term in the second line, we first transform $f'_m$ as $f'_m=f'_m/2+f'_m/2$ and then perform $m \leftrightarrow n$ on one of them.
Then, adding term (VI) to Eq.~\eqref{eq:III}, we obtain
\begin{align}
    (\mathrm{III})+(\mathrm{VI})&= \frac{1}{2} \biggl( f'_{nm} \frac{1}{\epsilon_{nm}} -\frac{f''_m}{2} -\frac{f'''_m}{6} \epsilon_{nm} \biggr)\mathrm{Re} \bigl[ v^b_n v^a_{nm}s^i_{mn} \bigr] -\frac{1}{24}(f''_m+f'''_m\epsilon_{nm} )\mathrm{Re}\bigl[ \partial_b\tilde{\epsilon}_{nm}v^a_{nm}s^i_{mn} \bigr]  \notag \\
    &=\biggl\{ \frac{f''_m}{4} \biggl( v^b_n-\frac{1}{6}\partial_b\tilde{\epsilon}_{nm} \biggr) +\frac{f''_m}{6} \biggl( v^b_n-\frac{1}{4}\partial_b\tilde{\epsilon}_{nm} \biggr) \epsilon_{nm} \biggr\}\mathrm{Re} \bigl[  v^a_{nm}s^i_{mn} \bigr] +\mathcal{O}(\epsilon^2_{nm}). \label{eq:III-IV}
\end{align}
Equations~\eqref{eq:II-IV-V} and \eqref{eq:III-IV} ensure the convergence of the two-band terms and can be used in numerical calculations around degenerate points.

\subsection{Multi-band terms} \label{subsec:multi-band}

On the other hand, careful treatment is required for the multi-band components in terms (I), (II), and (IV),
\begin{align}
    \bigl\{(\mathrm{I})+(\mathrm{II})+(\mathrm{IV}) \bigr\}_{\mathrm{multi\textendash band}}
    & = \biggl( -\frac{\mathcal{G}_{lm}}{2} \frac{1}{\epsilon_{ln}\epsilon_{nm}\epsilon_{lm}}+2\frac{\mathcal{G}_{nm}}{2}\frac{1}{\epsilon^2_{nm}\epsilon_{ln}}-\frac{f_n}{2} \frac{1}{\epsilon_{nm}\epsilon_{ln}}\biggr) \mathrm{Re}\bigl[ v^a_{nm}s^i_{ml}v^b_{ln}\bigr] \notag \\
    & = \frac{1}{2}\frac{1}{\epsilon_{nm}\epsilon_{ln}}\biggl( \frac{\mathcal{G}_{nm}}{\epsilon_{nm}}+\frac{\mathcal{G}_{ln}}{\epsilon_{ln}} -f_n-\frac{\mathcal{G}_{lm}}{\epsilon_{lm}} \biggr) \mathrm{Re}\bigl[ v^a_{nm}s^i_{ml}v^b_{ln}\bigr], \label{eq:multiband}
\end{align}
where we perform $l \leftrightarrow n$ in term (I) and $l \leftrightarrow m$ for the second term in the second line.
Equation~\eqref{eq:multiband} can be expanded with different infinitesimal quantities, depending on the type of degeneracy.
First, we consider the case $\epsilon_l=\epsilon_m$, where Eq.~\eqref{eq:multiband} becomes
\begin{align}
    (\mathrm{i})= \frac{1}{2}\frac{1}{\epsilon_{nm}\epsilon_{ln}}\biggl(  \frac{\mathcal{G}_{nm}}{\epsilon_{nm}}+\frac{\mathcal{G}_{ln}}{\epsilon_{ln}}-f_n-f_m-\frac{f'_m}{2}\epsilon_{lm}-\frac{f''_m}{6}\epsilon^2_{lm}-\frac{f'''_m}{24}\epsilon^3_{lm}\biggr) \mathrm{Re}\bigl[ v^a_{nm}s^i_{ml}v^b_{ln}\bigr]+\mathcal{O}(\epsilon^4_{lm}). \label{eq:multiband-(i)}
\end{align}
The convergence of this equation is guaranteed only in case (i) $\epsilon_n \neq \epsilon_m = \epsilon_l$.
In case (ii) $\epsilon_n=\epsilon_m=\epsilon_l$, Eq.~\eqref{eq:multiband-(i)} is further expanded as
\begin{align}
    (\mathrm{ii})&= \frac{1}{2}\frac{1}{\epsilon_{nm}\epsilon_{ln}}\biggl(  f_m+\frac{f'_m}{2}\epsilon_{nm}+\frac{f''_m}{6}\epsilon^2_{nm}+\frac{f'''_m}{24}\epsilon^3_{nm}+f_n+\frac{f'_n}{2}\epsilon_{ln}+\frac{f''_n}{6}\epsilon^2_{ln}+\frac{f'''_n}{24}\epsilon^3_{ln} \notag \\
    & \hspace{60pt}-f_n-f_m-\frac{f'_m}{2}\epsilon_{lm}-\frac{f''_m}{6}\epsilon^2_{lm}-\frac{f''_m}{24}\epsilon^3_{lm}\biggr) \mathrm{Re}\bigl[ v^a_{nm}s^i_{ml}v^b_{ln}\bigr]+\mathcal{O}(\epsilon^2_{nm},\epsilon^2_{ln}) \notag \\
    &=\frac{1}{2}\frac{1}{\epsilon_{nm}\epsilon_{ln}}\biggl(  \frac{f'_{nm}}{2}\epsilon_{ln}+\frac{f''_n}{6}\epsilon_{ln}(\epsilon_{lm}-\epsilon_{nm})-\frac{f''_m}{6}\epsilon_{ln}(\epsilon_{lm}+\epsilon_{nm}) \notag \\
    &\hspace{60pt} +\frac{f'''_n}{24}\epsilon_{ln}(\epsilon_{lm}-\epsilon_{nm})^2-\frac{f'''_m}{24}\epsilon_{ln}( \epsilon^2_{lm}+\epsilon_{lm}\epsilon_{nm}+\epsilon^2_{nm} )\biggr) \mathrm{Re}\bigl[ v^a_{nm}s^i_{ml}v^b_{ln}\bigr]+\mathcal{O}(\epsilon^2_{nm},\epsilon^2_{ln}) \notag \\
    &=\frac{1}{2}\frac{1}{\epsilon_{nm}\epsilon_{ln}}\biggl(  \frac{f'_{nm}}{2}\epsilon_{ln}+\frac{f''_{nm}}{6}\epsilon_{ln}\epsilon_{lm}-\frac{f''_{nm}}{6}\epsilon_{ln} \epsilon_{nm}-\frac{f''_m}{3}\epsilon_{ln} \epsilon_{nm}-\frac{f'''_m}{8}\epsilon_{ln}\epsilon_{nm}\epsilon_{lm} \notag \\ 
    &\hspace{60pt} -\frac{f'''_{nm}}{12}\epsilon_{ln}\epsilon_{nm}\epsilon_{lm}+\frac{f'''_{nm}}{24} \epsilon_{ln}\epsilon^2_{lm}+\frac{f'''_{nm}}{24} \epsilon_{ln}\epsilon^2_{nm} \biggr) \mathrm{Re}\bigl[ v^a_{nm}s^i_{ml}v^b_{ln}\bigr]+\mathcal{O}(\epsilon^2_{nm},\epsilon^2_{ln}) \notag \\
    &=\frac{1}{12}\biggl(  f''_m+\frac{f'''_m}{2}\epsilon_{nm}+\frac{f'''_m}{4}\epsilon_{lm}\biggr) \mathrm{Re}\bigl[ v^a_{nm}s^i_{ml}v^b_{ln}\bigr]+\mathcal{O}(\epsilon^2_{nm},\epsilon^2_{ln}).
    \label{eq:multiband-(ii)} 
\end{align}

Then, we proceed with the case $\epsilon_l \neq \epsilon_m$, using an expression equivalent to Eq.~\eqref{eq:multiband},
\begin{align}
    \frac{1}{2}\biggl( -f_n\frac{1}{\epsilon_{nm}\epsilon_{ln}}+ \mathcal{G}_{ln}\frac{1}{\epsilon_{lm}\epsilon^2_{ln}} +\mathcal{G}_{nm}\frac{1 }{\epsilon_{lm}\epsilon^2_{nm}} \biggr) \mathrm{Re}\bigl[ v^a_{nm}s^i_{ml}v^b_{ln}\bigr], \label{eq:multiband-(iii,iv)}
\end{align}
where we use $\mathcal{G}_{lm}=\mathcal{G}_{ln}+\mathcal{G}_{nm}$.
In this case, two types of degeneracy can be considered: (iii) $\epsilon_n=\epsilon_m$ and $\epsilon_l\neq\epsilon_n$ and (iv) $\epsilon_l=\epsilon_n$ and $\epsilon_n \neq \epsilon_m$, where Eq.~\eqref{eq:multiband-(iii,iv)} is each expanded as
\begin{align}
    (\mathrm{iii}) &=\frac{1}{2}\biggl(  \mathcal{G}_{ln}\frac{1}{\epsilon_{lm}\epsilon^2_{ln}} -f_n\frac{1}{\epsilon_{lm}\epsilon_{ln}} -\frac{f'_n}{2} \frac{1}{\epsilon_{lm}}+\frac{f''_n}{6}\frac{\epsilon_{nm}}{\epsilon_{lm}}  \biggr) \mathrm{Re}\bigl[ v^a_{nm}s^i_{ml}v^b_{ln}\bigr] +\mathcal{O}(\epsilon^2_{nm}), \label{eq:multiband-(iii)} \\
    (\mathrm{iv}) &= \frac{1}{2}\biggl(  \mathcal{G}_{nm}\frac{1}{\epsilon_{lm}\epsilon^2_{nm}} -f_n\frac{1}{\epsilon_{lm}\epsilon_{nm}} +\frac{f'_n}{2} \frac{1}{\epsilon_{lm}}+\frac{f''_n}{6}\frac{\epsilon_{ln}}{\epsilon_{lm}}  \biggr) \mathrm{Re}\bigl[ v^a_{nm}s^i_{ml}v^b_{ln}\bigr] +\mathcal{O}(\epsilon^2_{ln}). \label{eq:multiband-(iv)}
\end{align}
Equations~\eqref{eq:multiband-(i)}, \eqref{eq:multiband-(ii)}, \eqref{eq:multiband-(iii)}, and \eqref{eq:multiband-(iv)} guarantee convergence for cases (i), (ii), (iii), and (iv), respectively, and can be used in numerical calculations around degenerate points.

\end{widetext}
\bibliography{main.bib}

\begin{thebibliography}{62}
\bibitem [{not()}]{note1}%
  \BibitemOpen
  \bibinfo {title} {{The contribution from the degenerate points is included only in ``Total". This effective operation does not pose any problems because the contribution is sufficiently small.}}\BibitemShut {Stop}%
\end{thebibliography}
\end{document}